\documentclass[5p,times]{elsarticle}

\usepackage{amsmath}
\usepackage{subcaption}
\usepackage{booktabs}
\usepackage{textcomp}
\usepackage{lineno}
\usepackage{miller}
\PassOptionsToPackage{hyphens}{url}
\usepackage{hyperref}
\modulolinenumbers[5]

\newcommand{\textapprox}{\raisebox{0.5ex}{\texttildelow}}

\journal{Computational Materials Science}

\usepackage{hypernat}

\begin{document}
\frenchspacing

\begin{frontmatter}

\title{Application of artificial neural networks for rigid lattice kinetic Monte Carlo studies of Cu surface diffusion}

\author[helsinki]{Jyri Kimari\corref{correspondingauthor}}
\cortext[correspondingauthor]{Corresponding author}
\ead{jyri.kimari@helsinki.fi}
\author[helsinki]{Ville Jansson}
\author[helsinki,tartu]{Simon Vigonski}
\author[helsinki]{Ekaterina Baibuz}
\author[rio]{Roberto Domingos}
\author[tartu]{Vahur Zadin}
\author[helsinki]{Flyura Djurabekova}

\address[helsinki]{Helsinki Institute of Physics and Department of Physics, P.O. Box 43 (Pietari Kalmin katu 2), FI-00014 University of Helsinki, Finland}
\address[tartu]{Institute of Technology, University of Tartu, Nooruse 1, 50411 Tartu, Estonia}
\address[rio]{Instituto Polit\'ecnico de Nova Friburgo -- Universidade do Estado do Rio de Janeiro, Rua Sao Francisco Xavier, 524, 20550--900, Rio de Janeiro, RJ, Brazil}

\begin{abstract}

Kinetic Monte Carlo (KMC) is a powerful method for simulation of diffusion
processes in various systems. 
The accuracy of the method, however, relies on the extent of
details used for the parameterization of the model.
Migration barriers are often used to describe diffusion on atomic
scale, but the full set of these barriers may become easily
unmanageable in materials with increased chemical complexity or
a large number of defects.
This work is a feasibility study for applying a machine learning approach for Cu surface
diffusion. We train an artificial neural network on a subset of
the large set of $2^{26}$ barriers needed to correctly describe
the surface diffusion in Cu. Our KMC simulations using the
obtained barrier predictor show sufficient accuracy
in modelling processes on the low-index surfaces and display the correct
thermodynamical stability of these surfaces.

\end{abstract}


\end{frontmatter}

  \section{Introduction}
    
    Diffusion in crystalline material is an important phenomenon in many situations.
    For example, computational studies of irradiation damage may
    have a hard time finding agreement with experiments without accounting for bulk
    defect diffusion and annihilation after the initial collision cascade~\cite{yi2015direct}.
    Surface diffusion, on the other hand, is hypothesised to play a role
    in e.g. the events preceding vacuum arc breakdowns in devices with high electric field
    gradients~\cite{antoine2012erratum}. Vacuum arcs hinder the
    operation of many such devices, including particle accelerators, free-electron
    lasers and fusion reactors. The immediate motivation for studying the Cu
    surface specifically is the Compact Linear Collider (CLIC)~\cite{clic2016updated},
    which has been proposed to be built in CERN to succeed the Large Hadron Collider.
    
    Diffusion is difficult to study with molecular dynamics (MD) because it is
    a much slower process compared to the MD timestep, which has to be small enough to capture
    the atomic vibrations. Kinetic Monte Carlo (KMC) is a method suitable for longer time scale
    studies, such as simulations of the diffusion in crystalline solids. In this method, diffusion is approximated
    as a series of migration events (jumps) between potential energy minima. The
    jumps, while actually determined by Newton's equations of motion with
    atoms vibrating most of the time near potential energy minima, only occasionally
    crossing the energy barriers between them, can in aggregate
    be regarded as stochastic events that occur at rates determined by the height of those barriers.
    
    Such migration energy barriers must be known for each event that is to be considered in the KMC
    simulation. Depending on the desired specificity to distinguish between
    different events, the number of barriers that must be known may be too high to
    be calculated in a feasible time. Machine learning was proposed earlier as an
    alternative approach---the problem of too many barriers
    may be solved by only calculating accurately a subset of the barriers
    and obtaining the rest from a computationally inexpensive regression model.
    Djurabekova et al. used artificial neural networks~(ANN)
    to predict migration barriers in Fe-Cu bulk~\cite{djurabekova2007stability,djurabekova2007artificial}.
    Castin et al.~\cite{castin2008use,castin2009prediction,castin2009modelling,castin2010calculation,
    castin2011modeling,castin2012mobility,castin2014predicting,castin2017improved}, Pascuet et al.~\cite{pascuet2011stability}
    and Messina et al.~\cite{messina2017introducing} have since applied them to study bulk diffusion in
    various Fe-based alloys. Machine learning for surface diffusion
    barriers have been used at least by Sastry et al.~\cite{sastry2005genetic},
    who applied genetic programming for vacancy-assisted \hkl{100} surface migration barriers
    in Cu-Co alloy, and Verma et al.~\cite{verma2013cluster}, who built a cluster expansion model for
    Ag, Al, Cu, Ni, Pd and Pt \hkl(100) surface migration.
    
    In this paper, we will study the capabilities of ANNs to predict migration barriers on
    Cu surfaces. As an outcome of this study, we present a configuration of trained neural
    networks for parameterizing a rigid lattice KMC model of the (arbitrarily oriented) Cu surface.
    The networks and instructions to utilise them are published in the parallel Data in Brief
    publication~\cite{kimari2020data}.
    On the one hand, this work is an extension of the rigid lattice surface diffusion model
    developed earlier in our group~\cite{jansson2016long}, towards a more detailed description of the atomic
    environments and thus, hopefully, more accurate dynamics in the simulations.
    For the numerous challenges in the migration barrier calculations on
    the surface, we deploy solutions developed by Baibuz et al.~\cite{baibuz2018migration}.
    On the other hand, this work applies the methodology of
    refs.~\cite{djurabekova2007stability,djurabekova2007artificial,castin2008use,castin2009prediction,castin2009modelling,
    castin2010calculation,castin2011modeling,castin2012mobility,castin2014predicting,castin2017improved,
    pascuet2011stability,messina2017introducing} to a new system: the face-centered cubic (fcc) crystal surface.
    We will discuss various obstacles of the methods and how to overcome them.

    The KMC method, barrier calculations, and ANN models that we have used are described
    in section~\ref{sec:methods}. The results are presented in section~\ref{sec:results}
    and discussed in section~\ref{sec:discussions}. Conclusions about the
    applicability of ANNs for the considered problems are drawn in section~\ref{sec:conclusions}.

  \section{Methods}
  \label{sec:methods}
  
    \subsection{Kinetic Monte Carlo}
    \label{subsec:KMC}
      We used the residence-time KMC algorithm with the rigid lattice
      approximation, as implemented in the Kimocs code~\cite{jansson2016long}.
    
      KMC models the time evolution of a system
      by choosing events to be carried out one after another.
      The probability of an event to be chosen is proportional to its
      rate $\Gamma$, which in our model is given by the Arrhenius equation:
      \begin{equation}
        \label{eq:KMC_rate}
        \Gamma = \nu \exp\left(\frac{-E_\mathrm{m}}{k_\mathrm{B}T}\right),
      \end{equation}
      where $\nu$ is the attempt frequency, $E_\mathrm{m}$ is the migration
      energy barrier, $k_\mathrm{B}$ is the Boltzmann constant and $T$ is
      temperature. In our current model, $E_\mathrm{m}$ is strongly
      dependent on the local atomic environment of each jump, while
      $\nu$ is taken to be the same for all events.
      The $\nu$ parameter is connected to the vibration frequency of
      atoms, but it can also been seen as a scaling factor for the simulation time.
      
    \subsection{Barrier calculations}
    \label{subsec:barriers}
    
\begin{figure}
        \centering
        \includegraphics[width=0.5\linewidth]{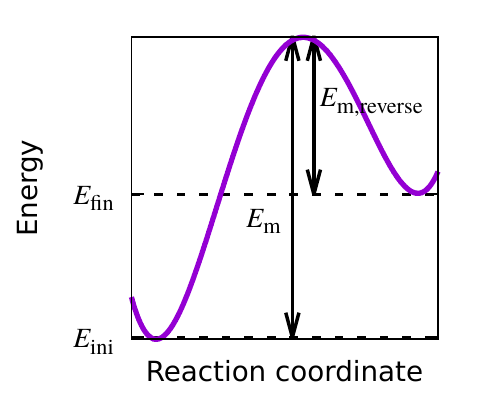}
        \caption{A schematic illustration of a 1-dimensional potential energy
                 surface from a NEB calculation. The barrier $E_\mathrm{m}$
                 and the reverse barrier $E_\mathrm{m,reverse}$ can be obtained
                 from the same calculation.}
        \label{fig:barrier}
      \end{figure}

      The barrier data set was calculated using the
      nudged elastic band (NEB)~\cite{mills1994quantum,mills1995reversible} method
      implemented in the LAMMPS molecular dynamics
      program~\cite{plimpton1995fast}.
      The NEB method finds the minimum energy path (MEP) between an initial
      and a final configuration. The barrier is the difference between
      the highest energy point along this path $E_\mathrm{max}$ (saddle point, since
      it is the maximum along the MEP and the minimum along other directions)
      and the initial state energy
      $E_\mathrm{i}$:
      \begin{equation}
        \label{eq:barrier}
        E_\mathrm{m} = E_\mathrm{max} - E_\mathrm{i}
      \end{equation}
      This is also illustrated in fig.~\ref{fig:barrier}.
      The potential energy function we chose was based on the molecular dynamics/Monte Carlo
      corrected effective medium (MD/MC-CEM) theory, developed by Stave et al.~\cite{stave1990corrected}.
      The potential is reported to describe the properties of the Cu surfaces well~\cite{sinnott1991corrected}.
      
      \begin{figure}
        \centering
        \begin{subfigure}{0.49\linewidth}
          \centering
          \begin{subfigure}{0.8\linewidth}
            \centering
            \includegraphics[width=\linewidth]{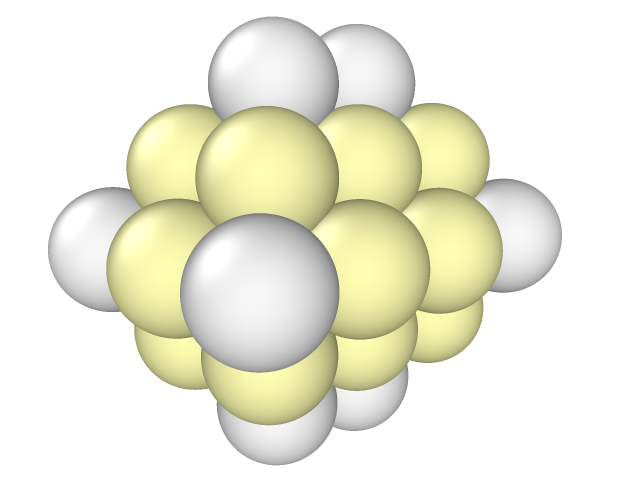}
          \end{subfigure}
          \begin{subfigure}{0.8\linewidth}
            \centering
            \includegraphics[width=\linewidth]{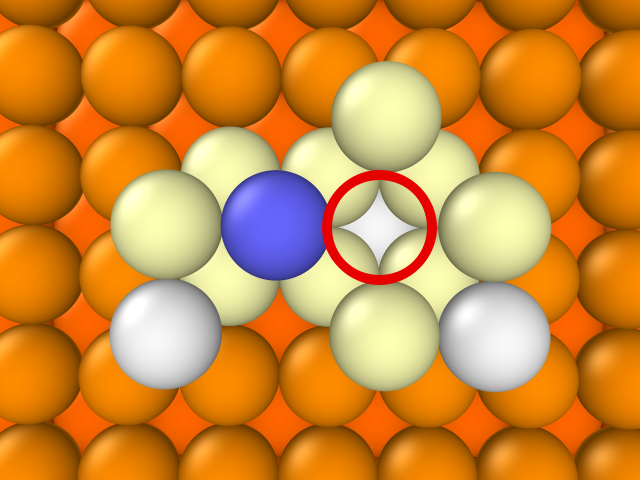}
          \end{subfigure}
        \end{subfigure}
        \begin{subfigure}{0.49\linewidth}
          \centering
          \includegraphics[width=0.7\linewidth, trim={5cm 0 5cm 0}, clip]{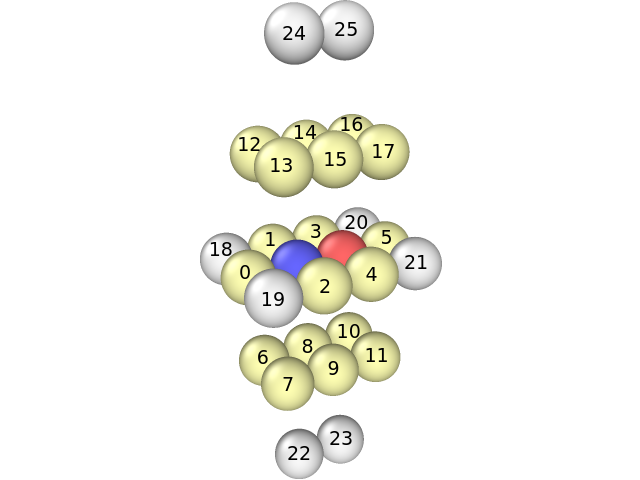}
        \end{subfigure}
        \caption{The local atomic environment (LAE) used in the process descriptor.
        	\emph{(Top left panel)} The octahedral 28-atom cluster containing the migrating
        	atom and its 1nn and 2nn sites. 1nn sites are coloured light yellow
        	and the 2nn sites are light grey.
        	\emph{(Right panel)} The indexing of the neighbour sites from 0 to 25 within the cluster.
        	The initial position of the migrating atom is marked blue
        	and the final position red.
        	\emph{(Bottom left panel)} Process ``1\,0\,0\,1\,1\,1\,1\,1\,1\,1\,1\,1\,0\,0\,0\,0\,0\,0\,0\,1\,0\,1\,1\,1\,0\,0'':
        	the blue atom jumps to the position circled in red.
        	The LAE has been embedded in the \hkl{100} surface.
                The figures were generated with Ovito~\cite{stukowski2009visualization}.}
        \label{fig:LAE}
      \end{figure}

      To connect the barriers to the corresponding
      processes during KMC simulations, a process descriptor is required.
      This same descriptor can also be used when training the ANN regressor.
      We describe the migration processes by their local atomic environments (LAE) before the jumps take place.
      We include in the LAE the first and the second nearest neighbours (1nn and 2nn) of the initial and
      the final position, as was done in the earlier parameterizations of Kimocs~\cite{jansson2016long}.
      In systems with the fcc structure, this definition of LAE covers 26 lattice sites in total (see fig.~\ref{fig:LAE}).
      We did not expand the LAE beyond the 2nn sites to keep the input space at a manageable size.
      The impact of the size of the local atomic environment on the accuracy of the surface migration barriers is subject to future study.
      
      In ref.~\cite{jansson2016long}, the LAE was described
      using a 4-dimensional vector $(a,\,b,\,c,\,d)$, where $a$ and $c$ are the number of 1nn atoms of the initial and
      the final position of the jump and $b$ and $d$ are the number of 2nn atoms of the initial and the final
      positions of the jump. We will hereafter refer to this approach as the ``4D description'',
      as it was named in ref.~\cite{baibuz2018migration}.
      KMC simulations that employ the 4D description have been found to produce good agreements with
      both MD and experimental results \cite{jansson2016long,zhao2016formation,baibuz2018migration,vigonski2017nanowire,jansson2019growth,jansson2020tungsten}.
      The 4D description does not include the
      information on the exact locations of the 26 LAE atoms,
      but only reflects the stability of the
      initial and the final positions by counting
      the number of neighbours in each neighbour shell.
      This descriptor is also not necessarily extensible
      to systems with multiple atomic species present.
      To save the location information,
      in our new descriptor the occupation state of each site is encoded with
      either~0~(vacant) or~1~(occupied).
      The descriptor is a 26-dimensional
      binary vector, referred to as the 26D description. The total number of
      different LAEs that can be distinguished with this descriptor is $2^{26}\approx 67$\,million.
      This choice of LAE encoding does not restrict the
      fcc surface orientation that can be described, and it is
      equally well applicable for bulk diffusion processes
      or systems with more than one atomic species, by
      denoting different elements with 1, 2, etc.
      This type of encoding is similar to what was used in
      refs.~\cite{castin2008use,castin2009prediction,castin2009modelling,castin2010calculation,
      castin2011modeling,castin2012mobility,pascuet2011stability,
      castin2017improved,messina2017introducing}
      to describe processes in bulk systems with the
      body-centered cubic (bcc) structure. A similar descriptor
      was also used by Trushin et al.~\cite{trushin2005self} for
      a \emph{self-learning} 2D KMC model fcc surfaces, and by Latz et al.~\cite{latz2012three}
      for the 3D expansion of that model.
      A self-learning KMC model, unlike a machine learning parameterized one, does not
      build a regression function for migration barriers, but simply calculates them on-the-fly and
      saves each result in a searchable catalogue for later use. This kind of model needs a descriptor for labelling
      the catalogue, whereas a machine learning model uses the descriptor as a way to encode input for the
      regression function.
      
      The 26D descriptor is not compact in the sense that some LAEs are
      mirror images of each other, and thus have the same migration
      energy barrier: multiple descriptors have the same expected output.
      Up to four different LAEs may belong into these ``families'' of
      equivalent processes.
      This is problematic from the machine learning point of view:
      even though the redundant symmetric cases may be included in the
      training data for the ANN, training will never be perfect and
      the network output will be different for each case.
      If the ANN predicts different barriers for mirrored, but otherwise physically
      equivalent, LAEs, diffusion may work differently e.g. on different \hkl{100}
      facets like (100) and (010), or be biased towards some directions. This problem was
      solved by removing the redundant processes from the training set
      systematically: only one process represents each family in
      the training set. When calling the network to produce the
      barrier of a given jump, the LAE is first transformed to
      correspond to the ``representative'' that is shown (or would be 
      shown, in the case of extrapolation) to the network
      in the training.
      
      In the remainder of this paper we will refer to the
      LAE configuration around the initial and final jump
      positions up to the second nearest neighbour shells as
      the LAE cluster. Even though, as it is said, the 26D
      LAE cluster contains the information about the
      neighbours up to the second nearest neighbour shell,
      for barrier calculations by NEB, we embed such a
      cluster in a larger lattice, where all the sites are occupied
      by the same atoms and affect the calculations in
      the systematic manner---calculating the migration barriers
      within an isolated cluster of a few dozen atoms would not
      be realistic. As a first approach we embed the LAE cluster in Cu
      bulk for the NEB calculations, similarly as we did in
      refs.~\cite{jansson2016long,baibuz2018migration}.
      This is valid especially in studies
      where only a limited number of vacancies is present in the LAE.
      However, when calculating surface migration barriers, the LAE
      is on average half-empty. Previously we discovered~\cite{baibuz2018migration}
      that in some cases embedding such an LAE, essentially a large
      vacancy cluster (void),
      behaves differently inside bulk compared to the surface, causing
      very strong forces and unphysically high barriers in the
      range of tens of eV. We thus returned to the original recipe,
      calculating all the barriers  with the LAE clusters embedded in
      a surface (see fig.~\ref{fig:LAE}).
      
      Embedding the cluster in a surface is more complicated than in bulk.
      It is not immediately obvious which surface should be used, and how each LAE should
      be oriented to be best fit into the surface. To answer these questions, we developed
      an automated procedure, where
      \begin{enumerate}
        \item Three different surfaces were considered: \hkl{100}, \hkl{110} and \hkl{111}.
              Within the local environment limited to 2nn sites, also the higher index surfaces will
              resemble one of these lowest-index cases, and thus the resulting model
              can be generally applied on any surface orientation.
        \item For each of the three surface orientations, trial configurations were generated by
              embedding the LAE cluster in every possible orientation (with the constraint that the
              lattice points of the LAE have to match the lattice points of the surface).
              The LAE was embedded deep enough that at most one layer of atoms
              was above the surface of the surrounding lattice.
        \item The centre-of-mass of the LAE cluster was calculated in each of the
              trial configurations. The configuration with the lowest
              centre-of-mass with respect to the surface was considered to
              be the most stable one, and was selected to be used in the
              NEB calculation to find the barrier in this LAE.
      \end{enumerate}
      We used a simulation cell of
      approximately $45\times45\times25\,\AA^3$ in
      size. Depending on the specific crystal orientation, the exact dimensions of the cell varied.
      Expanding the cell beyond this size did not affect the barrier value significantly.
      The boundary conditions were periodic in the horizontal dimensions and
      two fixed atom layers at the bottom.
      
      Eleven replicas were used in the NEB calculations. In addition to the force given by the
      MD/MC-CEM potential, a tethering spring force was applied on each atom in the same way as in \cite{baibuz2018migration}. 
      The tethering force keeps the atoms close to their initial lattice sites, but 
      does not directly contribute to the potential energy of the system.
      Previous KMC simulations with barrier sets for Cu and Au, calculated using the tethering
      method, have also shown good agreements with both MD and experimental results \cite{baibuz2018migration,vigonski2017nanowire,jansson2019growth,jansson2020tungsten}.
      
      In our case, if any atom slips to unintended lattice sites during the NEB relaxation,
      the process will no longer correspond to its 26D descriptor
      and thus the obtained descriptor-barrier pair cannot be used
      in the KMC simulations. This is a challenge specific to
      the rigid lattice KMC.
      The problem is more severe when performing NEB calculations
      on the surface, where there are other adatoms
      around the jumping atom, which may also have very
      few neighbours and thus are not strongly bound to the
      their lattice site. These
      loosely bound adatoms tend to easily ``follow'' the jumping
      atom, or otherwise leave their original lattice sites.
      The tethering force constant, as implemented in LAMMPS, was set to 2.0\,eV/\AA$^2$. For details
      about the tethering force approach, see~\cite{baibuz2018migration}.

      The first set of barriers was calculated for adatom migration processes
      on flat \hkl{100}, \hkl{110} and \hkl{111} surfaces, with the migrating atom
      surrounded by different configurations of atoms in the same atomic layer.
      The set formed by these LAEs will be referred to as \emph{flat surface processes}
      for the remainder of this paper.
      The ANNs were first tested by training and predicting barriers
      within this flat surface set, comprising 3168 processes.
      
      The set was then expanded by calculating the barriers for
      all processes where every atom within the LAE cluster and the migrating
      atom had at least three 1nn atoms around it when embedded
      on a surface. In total, 11\,652\,085
      processes were found in this category---approximately 17\,\%
      of the entire LAE space.
      Attempting to calculate the barriers for the rest of the
      configurations, where some atoms have less than three neighbours,
      is likely to produce artificial results---even with the tethering 
      force applied. In KMC simulations with
      only 1nn jumps permitted, it is still beneficial to include these
      unstable events in the simulation as they can serve as intermediate steps
      in e.g. 2nn or 3nn jumps that would be possible in a real system.
      An ANN regression model, that has learnt the general tendencies of the
      LAE dependent migration barriers, is likely to give more realistic
      values for the barriers of these events than NEB calculations that are less
      than reliable far from potential energy minima.
      Thus, ANNs are a good way to expand the NEB-calculated set of barriers
      to allow for more realistic kinetics.
      
    \subsection{Artificial neural networks}
    \label{subsec:ANN}
    
      ANNs are a class of machine learning
      methods that can be used for classification and function regression.
      A regressor is trained with known input-output pairs
      to learn the underlying function and also to predict the output for previously unseen input.
      In this work, ANNs were fitted to predict values of the migration barrier
      function $E_\mathrm{m}=E_\mathrm{m}(\mathrm{LAE})$.
      
      Two different ANN models were studied: multilayer perceptrons (MLP)
      and radial basis function (RBF) networks. The MLP implementation was
      taken from the Fast Artificial Neural Network (FANN) library~\cite{nissen03},
      and the RBF networks are from Python's SciPy package~\cite{scipy}.
      
      \subsubsection{Multilayer perceptrons}
      
\begin{figure}
          \centering
          \includegraphics[trim={0 0.5cm 0 0.4cm}, clip, width=0.8\linewidth]{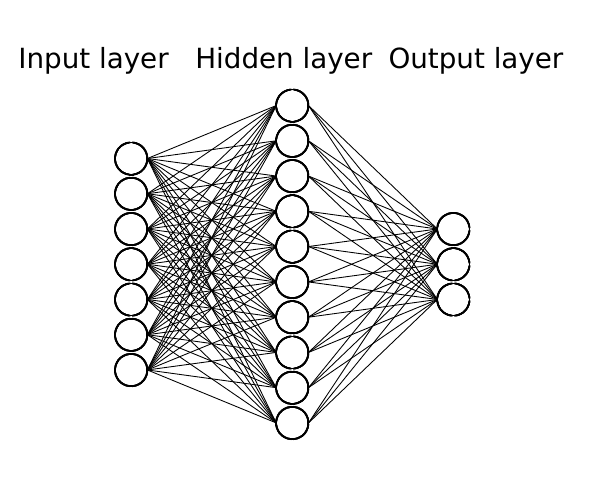}
          \caption{A schematic illustration of a fully connected feed-forward MLP with
          	7 input nodes, 10 hidden nodes and 3 output nodes.}
          \label{fig:ANN}
        \end{figure}

        MLPs consist of one input layer, one output layer, and one or more hidden
        layers between them. The layers hold nodes that are connected to
        each other. In a fully connected feed-forward network without
	shortcut connections, all nodes of each layer are connected to
        all nodes of the next layer (see fig.~\ref{fig:ANN}
        for a schematic illustration). This was one of the network
        structures used in this work. The other type of MLP structure
        used here was a cascade network where hidden nodes are added one
        by one during training so that each new node is connected to each of
        the old nodes. This way the network will have $N$ hidden nodes
        in $N$ hidden layers with shortcut connections between each layer.
        
        The input is mapped to the output by passing it through
        the hidden nodes: each node calculates the weighted sum of
        its inputs and passes this value to the next nodes through an activation function.
        The weights are chosen in an iterative training
        process to produce minimal error in the training set.
        The training algorithm used
        in this work was the FANN library implementation of the
        improved resilient back-propagation (iRPROP) algorithm
        proposed by Igel and H\"usken~\cite{igel2000improving}.
        
        The purpose of the activation function is to allow for
        non-linear regression. For the MLPs with static layout,
        we used sigmoid (logistic) activation functions in the hidden
        nodes. In the cascade networks the Cascade2~\cite{fahlman1990cascade}
        algorithm sets the hidden node activation functions automatically.
        In the output node we found linear activations function to perform best
        in all MLP networks. With the default sigmoid activation function (scaled to accommodate
        for the full range of energy barrier values)
        in the output node, the training would often stagnate, with the network
        weights drifting towards extreme values before sufficient accuracy was
        achieved. Cascade networks would completely diverge for this problem
        with a sigmoid output activation function.
        
        To avoid using negative barrier values
        given by the linear activation, we take the barrier to
        be $E_\mathrm{m}=\max(0,\,E_\mathrm{ANN})$
        during the simulations.
        
        Another post-processing step was to forbid processes that
        would lead to complete detachment of atoms from the substrate.
        Examples of such processes are not included in the training set,
        so the ANN-predicted barriers for them are greatly underestimated.
        
        The number of input and output nodes is determined by dimensionality of the problem
        at hand. In the case of learning the 1D
        barrier values that correspond to the 26D input, the
        input layer will have 26 nodes, and the output layer will have
        one node. The number of hidden layers and nodes is another matter
        of optimisation. In this study, a single hidden layer with 35
        nodes was found to be optimal for the static (non-cascade) network.
        For the cascade networks, 30--70 nodes were needed for
        reaching a sufficient accuracy. Overfitting was not observed: the root
        mean square error in the validation set was equal to the error in the
        training set.

      \subsubsection{Radial basis function networks}
      
        RBF networks are somewhat similar to MLPs. The difference is that the activation
        functions are radially symmetric functions that depend on the distance $r$ of the
        input vector of the \emph{prototype vector} of each node. The Gaussian function
        is an example of a radial basis function. In this work,
        we used multiquadric functions
        \begin{equation}
          f(r) = \sqrt{\left(\frac{r}{\varepsilon}\right)^2+1}
        \end{equation}
        as basis functions; $\varepsilon$ is a width parameter that was
        automatically adjusted by the library function.
        
        RBF networks usually have only one hidden layer, with one basis function
        corresponding to each hidden node. The nodes have their own prototype
        vectors, and there is one weight vector that has a dimensionality
        equal to the number of hidden nodes. The weights are set by a matrix
        inversion that is faster than the iterative training procedure
        used with MLPs, but very memory-intensive for large training sets.
        For training sets beyond the~\textapprox3000 barrier flat surface
        set, this could not be done with the available resources---only
        MLP networks were trained with the full 11.7 million barrier data set.
        
        The prototype vectors for the RBF networks can be chosen from among the input vectors
        in the training set and the number of these vectors could in principle
        be smaller than the total number of data points. In the SciPy
        library this functionality is not implemented---instead, there
        is always one prototype vector for each data point. As a result,
        the accuracy of the obtained RBF network will always be perfect
        for the training set. For this reason, in this work, only 50\,\% of the 3000 barrier data set
        was used to train the RBF networks in order to get an estimate
        on how well they can actually learn the migration energy function. To make
        a fair comparison to MLPs, these were also trained using only 50\,\% of the flat surface
        data.
        
      \subsubsection{Error reduction techniques}
      \label{subsubsec:additional}
        
        When the barrier set was expanded from the initial~\textapprox3000
        barriers to the 11.7 million barriers, the MLP accuracy decreased
        significantly, with the cascade networks performing the best.
        Two additional techniques were used to reduce the error in barrier prediction.
        Firstly, the barriers calculated on each surface---\hkl{100}, \hkl{110}, or
        \hkl{111}---were treated as separate sets and different ANN predictors
        were trained to predict barriers on each of these sets.
        The issue of knowing which predictor to call to get the barrier for an arbitrary LAE
        encountered during KMC was solved by introducing an ANN classifier.
        This network uses the same LAE input encoding, but instead of
        energy output, it outputs the \emph{class}, or the surface that
        the input LAE corresponds to. We used 1-of-C encoding for the
        surfaces, meaning that the classifier has three output nodes
        with each node corresponding to one of the surface classes. A very
        good success rate was achieved for the classifier (see table~\ref{tab:confusion}).
        During the KMC simulation, every encountered LAE is first
        passed to the classifier and an appropriate regressor based on the classifier
        output is used to obtain the migration barrier. In the case that more than one classifier
        output signals non-zero value (the classifier is uncertain of the correct surface orientation),
        the barrier will be calculated as a weighted average of multiple regressors.
        \begin{table}
          \centering
          \caption{Confusion matrix for the surface classifier ANN.
                   The numbers on the diagonal represent correctly classified
                   LAEs. The total success rate was 99.39\,\%.}
          \label{tab:confusion}
          \begin{tabular}{crrr}
            \toprule
                    &      \multicolumn{3}{c}{Classified as:} \\
            \cline{2-4}
            Actual: &     \hkl{100} &     \hkl{110} &     \hkl{111} \\
            \midrule
            \hkl{100} & 1\,115\,537 &        6640 &     17\,104 \\
            \hkl{110} &        3310 & 1\,470\,607 &     21\,242 \\
            \hkl{111} &        4580 &     18\,582 & 8\,994\,483 \\
            \bottomrule
          \end{tabular}
        \end{table}
        Using this technique, the RMS prediction error decreased by approx. 10\,\%
        compared to using only a single predictor for the entire data set.

        Secondly, groups of ANN regressors that were trained on
        different, random subsets of the training data
        were combined into regressor ensembles.
        The predicted barrier for each LAE was taken to be the
        average of the predictions given by the individual networks.
        The individual regressors
        were originally the products of 5-fold cross-validation, but
        combining them into ensembles turned out to make better
        predictions than any of the components.
        The RMS prediction error decreased by 13--19\,\%
        compared to using only a single predictor for each surface.
        Combining the ensembles and the classifier reduced the RMS error
        by a total of 21.8\,\%: from 0.110\,eV to 0.086\,eV.
        
      \subsubsection{Setup of the kinetic Monte Carlo simulations}
      \label{subsubsec:KMC}
        
        In addition to calculating the RMS error estimate, the accuracy
        of the barrier predictions was assessed by
        implementing the ANN barrier function into the Kimocs code
        and running KMC simulations with the predicted barriers. Three different
        simulation scenarios were considered:
        \begin{enumerate}
          \item Flattening of a 12 monolayer (576 atoms) cuboid nanotip
                on the three lowest-index fcc surfaces, \hkl{100}, \hkl{110}, and \hkl{111}.
                The system dimensions are the same as the test
                cases in ref.~\cite{jansson2016long}.
          \item Relaxation of nanoparticles with different initial shapes.
                This will show how well the model captures the correct thermodynamics---every
                shape should relax approximately to the shape of the Wulff construction. Each
                shape was relaxed in three different sizes:~\textapprox900 atoms,~\textapprox1400 atoms,
                and~\textapprox2100 atoms. The Wulff constructions were created with the Atomic Simulation Environment~\cite{ase-paper},
                using the surface energies reported by the authors of the Cu interatomic potential~\cite{sinnott1991corrected}.
          \item Stability of \hkl<110> nanowires.
                Vigonski et al. found crossing Au nanowires to break through surface diffusion
                in KMC simulations near the junction points of the wires~\cite{vigonski2017nanowire}.
                Cu has the same lattice structure as Au, and thus it is interesting whether
                crossing Cu nanowires are found to behave similarly. Cu nanowire networks
                have applications in the large-scale fabrication of transparent and flexible
                electronics~\cite{mallikarjuna2016photonic}. The \hkl<110>
                orientation was chosen since wires oriented this way can be constructed to
                have the thermodynamically favourable \hkl{111} surfaces on all sides.
                Thick (1.1\,nm in radius) \hkl<110> Cu wires were found to be indefinitely
                stable in the original 4D parameterization of Kimocs~\cite{jansson2016long},
                so the radius was reduced down to~\textapprox 0.5\,nm to observe fragmentation.
                Both single and crossing wires were simulated.
        \end{enumerate}
      
  \section{Results}
  \label{sec:results}
    
    \subsection{Regression accuracy}
    \label{subsec:accuracyresults}
    
\begin{figure}
      \centering
      \includegraphics[width=\linewidth]{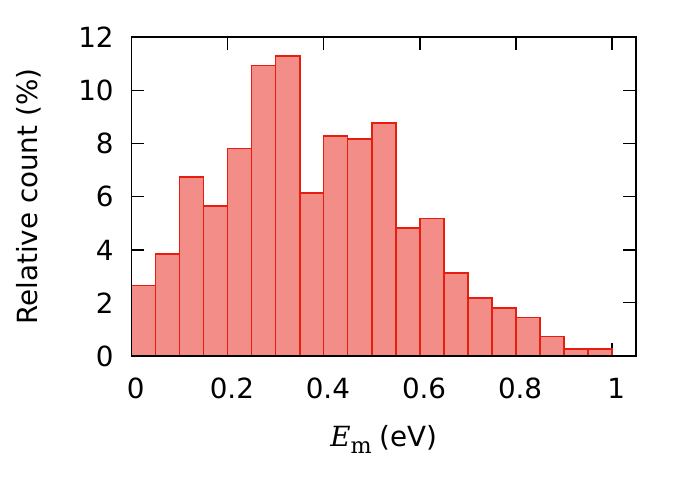}
      \caption{Distribution of migration barriers of the 3168 flat surface
               processes.}
      \label{fig:flat_histogram}
    \end{figure}

\begin{figure*}
      \centering
      \begin{subfigure}{0.32\linewidth}
        \centering
        \includegraphics[width=\linewidth]{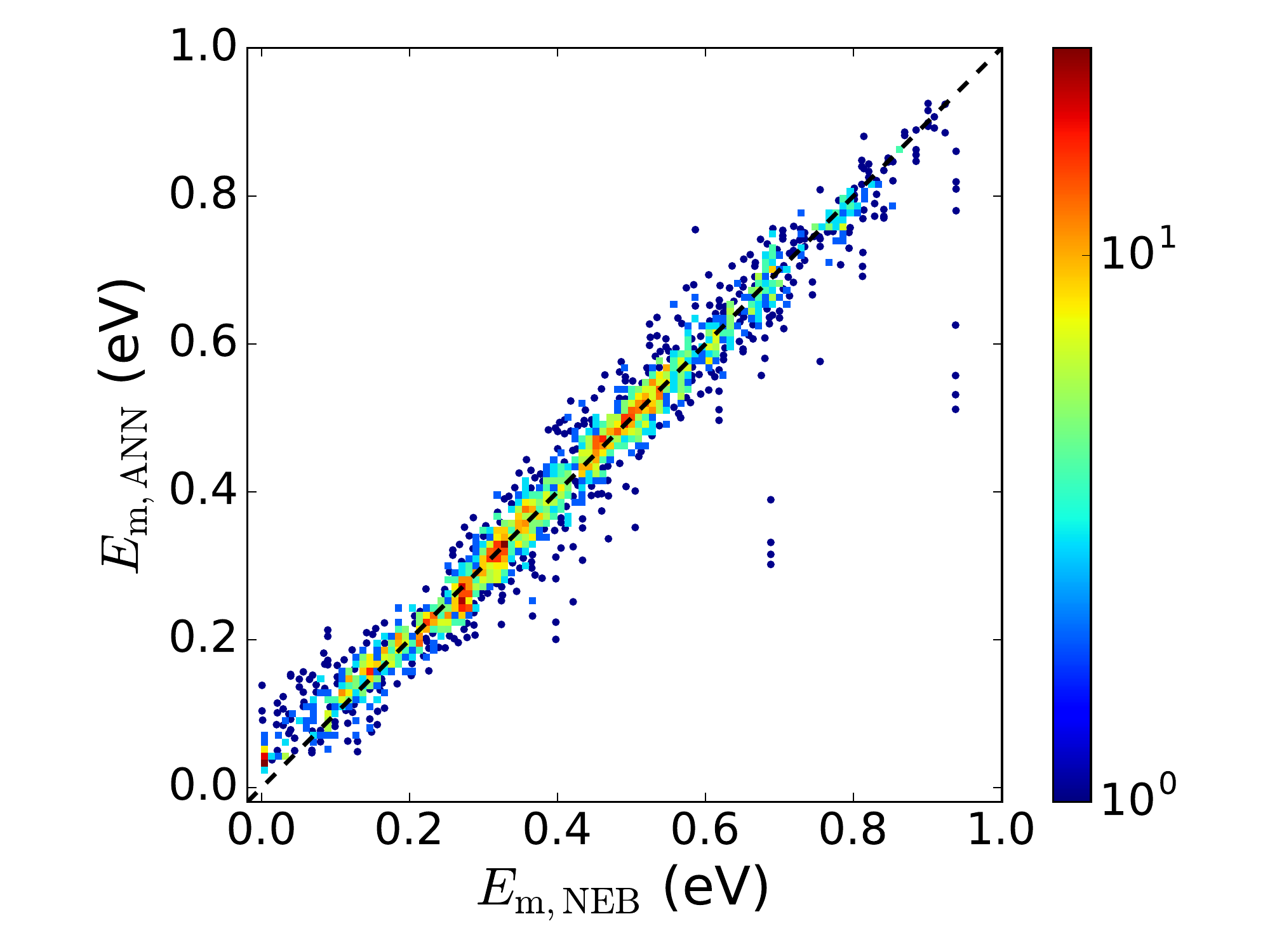}
      \end{subfigure}
      \begin{subfigure}{0.32\linewidth}
        \centering
        \includegraphics[width=\linewidth]{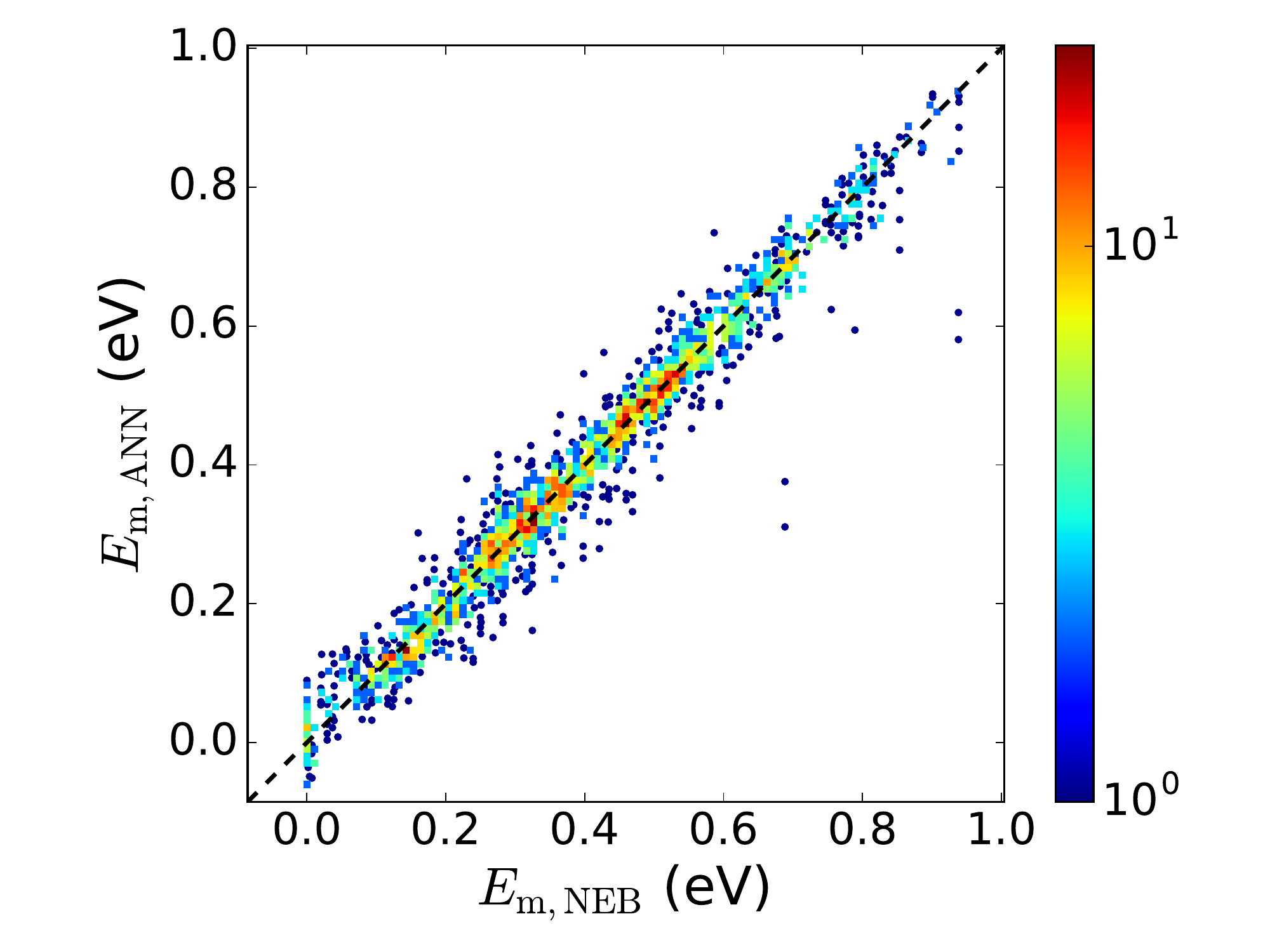}
      \end{subfigure}
      \begin{subfigure}{0.32\linewidth}
        \centering
        \includegraphics[width=\linewidth]{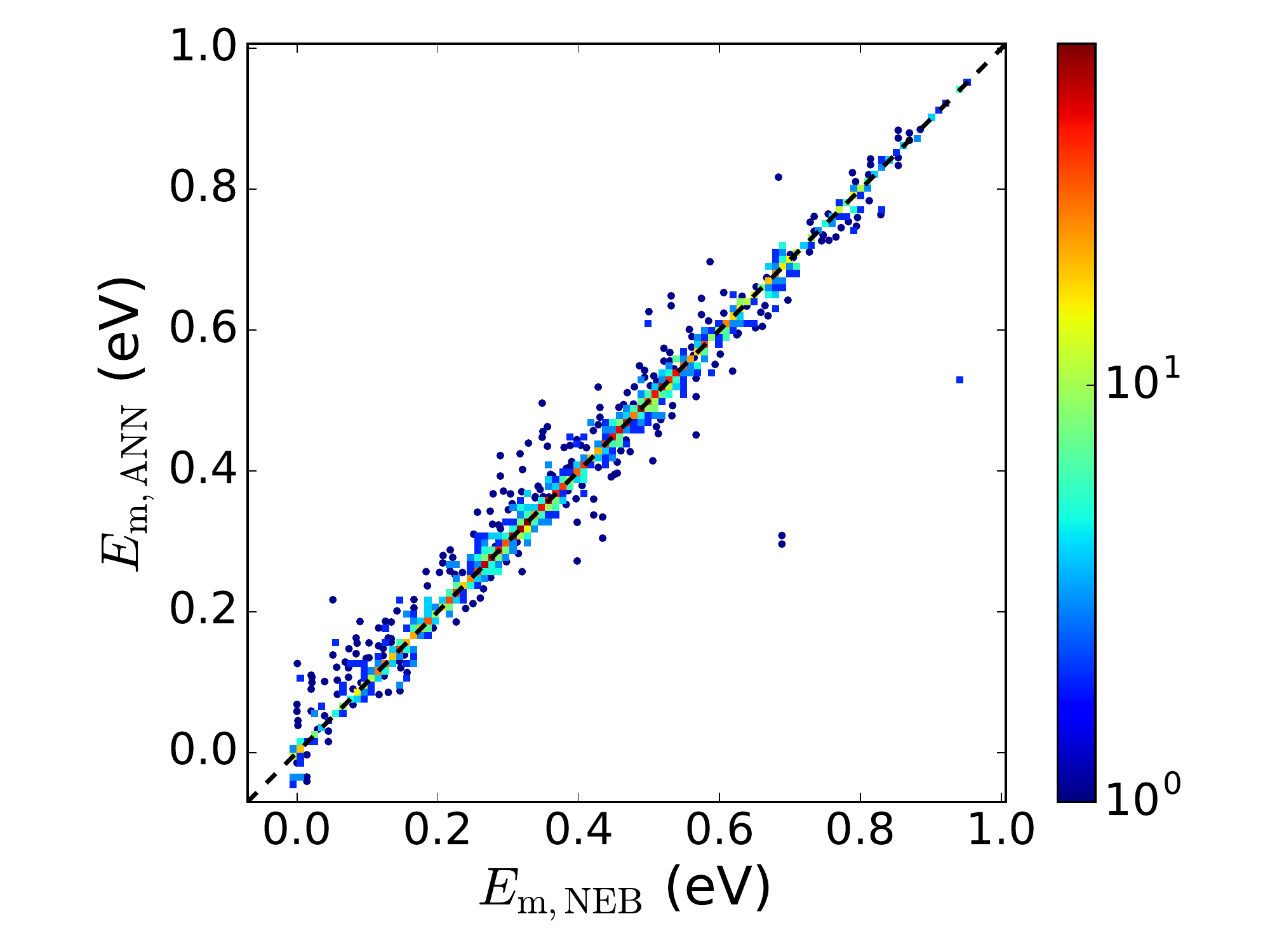}
      \end{subfigure}
      \caption{The static MLP \emph{(left)}, the
                cascade MLP \emph{(centre)}, and the RBF networks \emph{(right)}
                accuracy for flat surface migration barrier prediction.
      		All of the networks were trained using 50\,\% of the data set.
      		Colouring is according to the point density on a logarithmic scale.
      		The RMS errors were 0.036\,eV (static MLP), 0.033\,eV (cascade MLP)
      		and 0.024\,eV (RBF).}
      \label{fig:correlation_flat}
    \end{figure*}

    Fig.~\ref{fig:flat_histogram} shows the distribution of barriers
    in the initial set of~\textapprox3000 flat surface processes.
    Fig.~\ref{fig:correlation_flat} shows the accuracy
    of the static MLP, cascade MLP and RBF networks in this flat surface
    set. 50\,\% of the set was used in training, while the correlation is plotted
    in the full flat surface set.

\begin{figure}
      \centering
      \includegraphics[width=\linewidth]{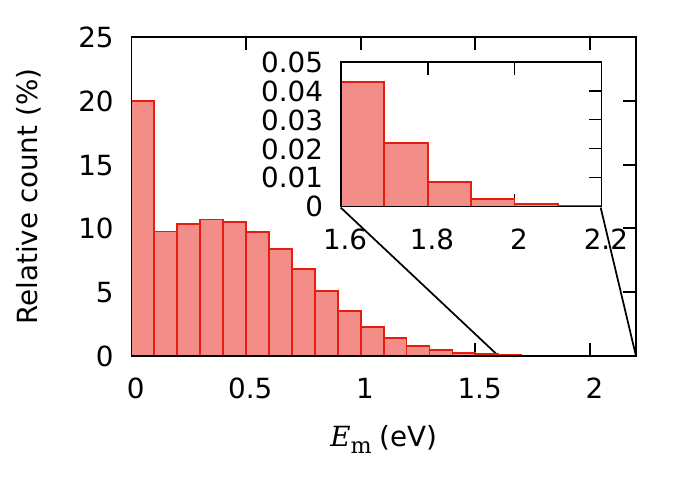}
      \caption{Distribution of migration barriers in the full~\textapprox11.7
               million barrier data set.}
      \label{fig:full_histogram}
    \end{figure}

    The migration energy distribution in the full~\textapprox11.7 million
    barrier set is shown in fig.~\ref{fig:full_histogram}. The large quantity
    of 0\,eV barriers are for processes that are spontaneous either in one
    direction (e.g. jumping towards a much higher number of 1nn atoms), or, less frequently, both
    directions. The latter processes include some events on the
    \hkl{111} surface that happen via hexagonal close-packed (hcp) sites.
    The full barrier set can be found as a Data in Brief entry~\cite{kimari2020data}.
    
\begin{figure}
      \centering
      \includegraphics[width=\linewidth]{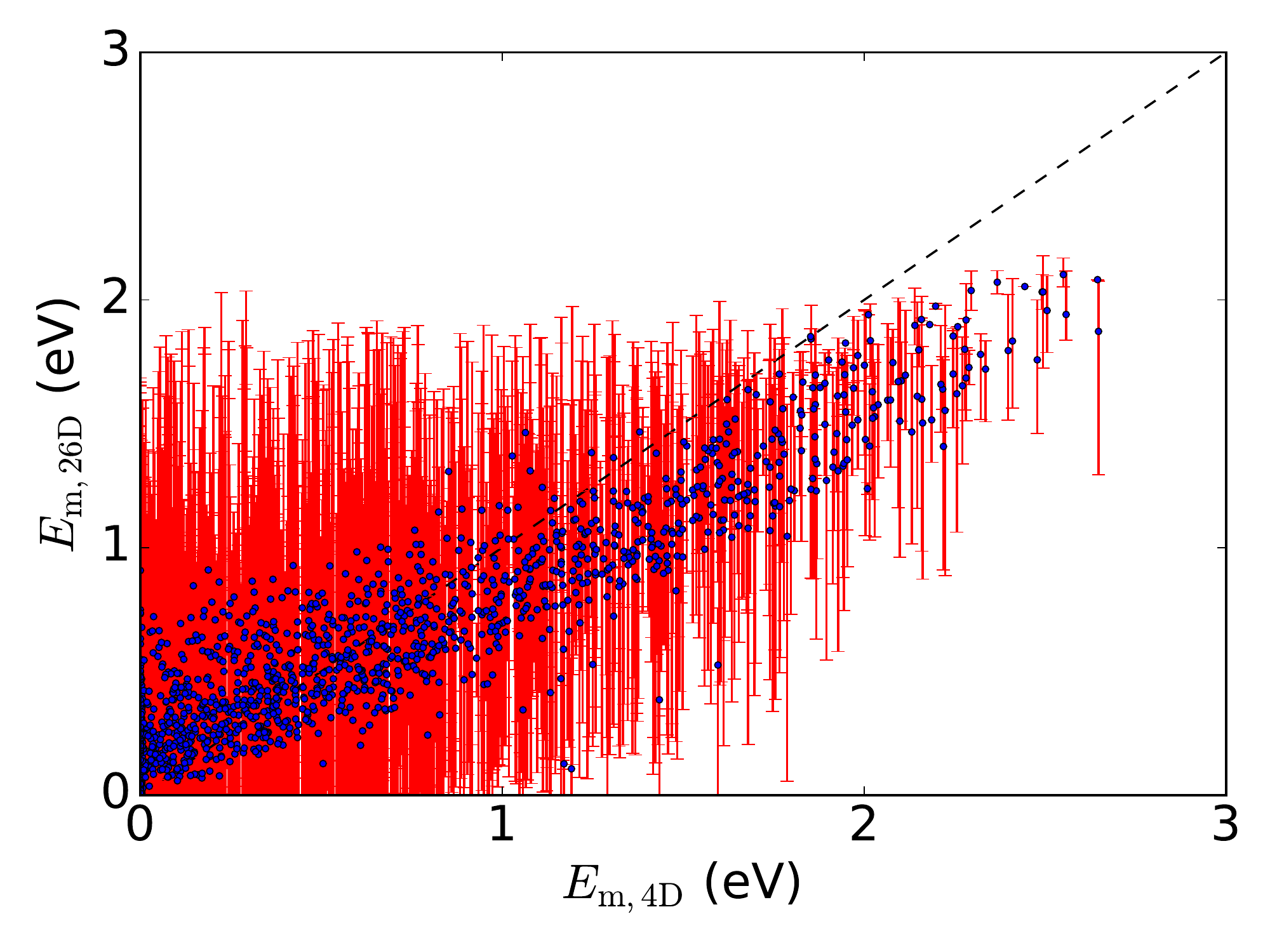}
      \caption{Comparison of barriers of the full barrier set to
      		the 4D-parameterized \emph{Cu set 1} of ref.~\cite{baibuz2018migration,baibuz2018data}
      		that was calculated without the tethering force.
      		Correspondence between the 26D and the 4D descriptions is
      		many-to-one; red bars show the minimum and the maximum values,
      		which are calculated for all possible permutations of the same 4D
      		description. The blue dots are the mean values of the barriers in
      		that range. Barriers over 1\,eV are generally lower in the 26D set than in
      	 	the 4D set.}
      \label{fig:compare_notether}
    \end{figure}

\begin{figure}
      \centering
      \includegraphics[width=\linewidth]{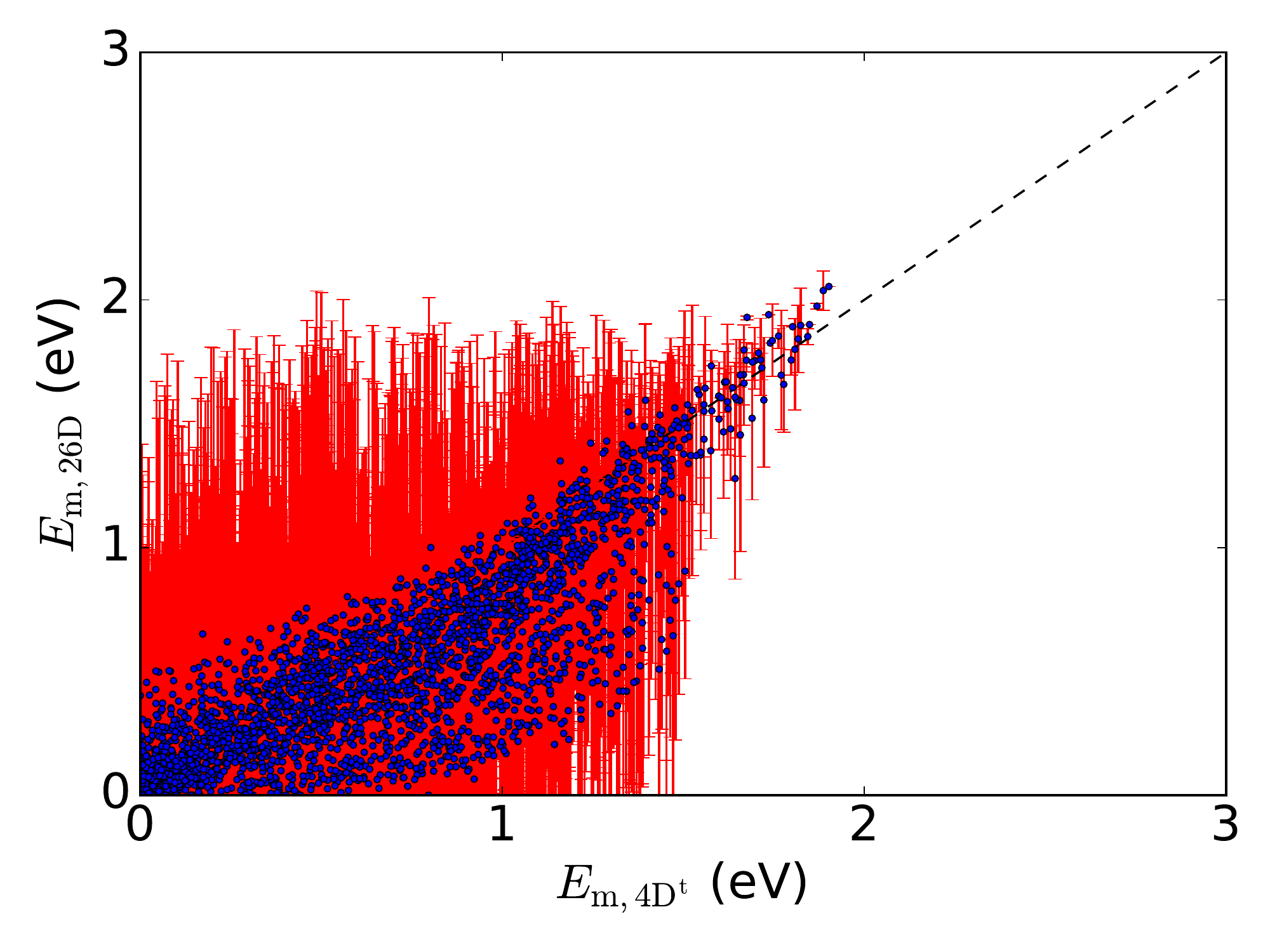}
      \caption{Comparison of barriers of the full barrier set to
      		the 4D-parameterized \emph{Cu set 2} of ref.~\cite{baibuz2018migration,baibuz2018data}
      		that was calculated with the additional tethering force.
      		Red bars and the blue dots are the same as in fig.~\ref{fig:compare_notether}.
      		Most of the mean values correlate well with the values of \emph{Cu set 2}, but there is
      		a region where the 26D set estimates barriers to be lower than the
      		corresponding 4D barriers.}
      \label{fig:compare_tether}
    \end{figure}

    Comparisons between the 11.7 million barrier full set
    and the 4D-parameterized barrier sets
    are shown in figs.~\ref{fig:compare_notether} and~\ref{fig:compare_tether}.
    In these figures the error bars indicate the highest and the lowest values
    of the barriers, which were calculated for different permutations within
    the same 4D description. The dark dot in the middle shows the value of the
    mean barrier calculated for all permutations. The mean value is calculated as a simple arithmetic average.
    Following the notation used in refs.~\cite{baibuz2018migration,baibuz2018data}, we will refer to two of such
    sets as \emph{Cu set 1} and \emph{Cu set 2}.
    Correspondence between physical local atomic environments and the 4D descriptors
    is not one-to-one, and thus the 4D barrier sets have been constructed by
    selecting one representative case from the family of environments
    that correspond to each 4D descriptor. \emph{Cu set 1} was constructed by randomly
    selecting one of the permutations of nearest neighbors, while \emph{Cu set 2}
    was selected by determining the lowest energy configurations of the initial and final sites of a jump.
    The tethering force approach was applied to the barrier calculations in \emph{Cu set 2}, with the same 
    tethering force constant 2.0\,eV/\AA$^2$ as in this work.
    The general trend of barriers in our 26D set is similar to the 4D sets, especially to
    \emph{Cu set 2}. It can be seen that the groups of configurations that correspond to the same 4D descriptor
    can have deviations of a few eV in energy barrier between themselves.

\begin{figure}
      \centering
      \includegraphics[width=\linewidth]{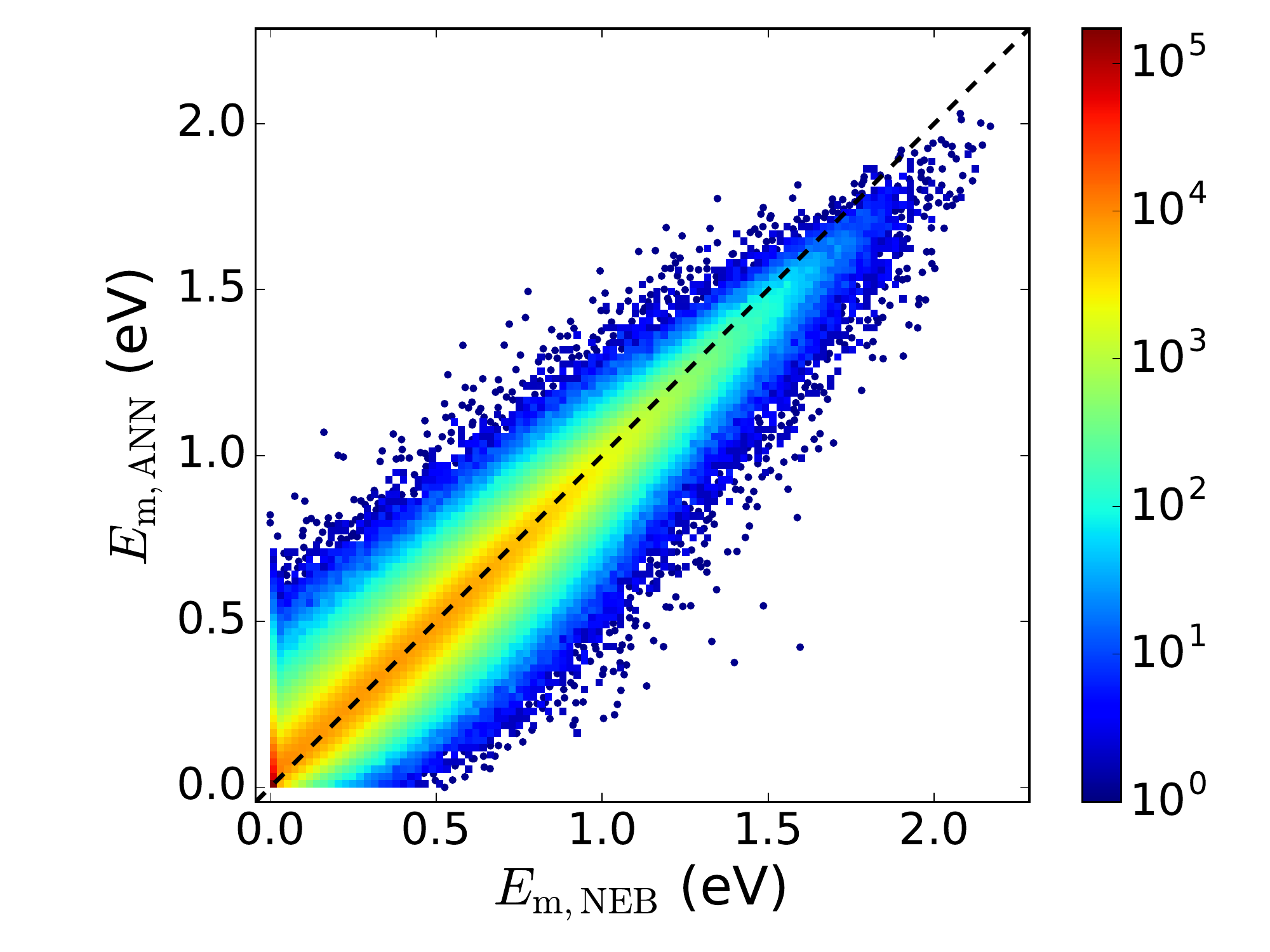}
      \caption{Accuracy of the combination MLP in the full~\textapprox11.7 million barrier data set.
      		The plotted result is from 16 individual networks: an ensemble
      		of five for each surface type (\hkl{100}, \hkl{110} and \hkl{111}) and
      		a classifier to combine the ensembles. The RMS error is 0.086\,eV.}
      \label{fig:FANN_full}
    \end{figure}

    The combination MLP regressor accuracy in the full~\textapprox11.7 million
    barrier set is shown in fig.~\ref{fig:FANN_full}. As explained in
    section~\ref{subsubsec:additional}, the final barrier predictor
    consists of three ensembles (one for each low-index surface) of five regressors each
    and a classifier to combine them. Training specialist networks to predict migration barriers
    on different physical surfaces improved total regression accuracy by 10\,\%, compared to training a single regressor
    to the entire data set. We propose that this improvement can be explained by
    the networks implicitly learning the effect of different surface relaxation
    and surface stress present on the different surfaces. These surface effects affect the energy of
    all surface atoms in the system, and thus modify the surface migration energy barriers.
    
\begin{figure}
      \centering
      \includegraphics[width=\linewidth]{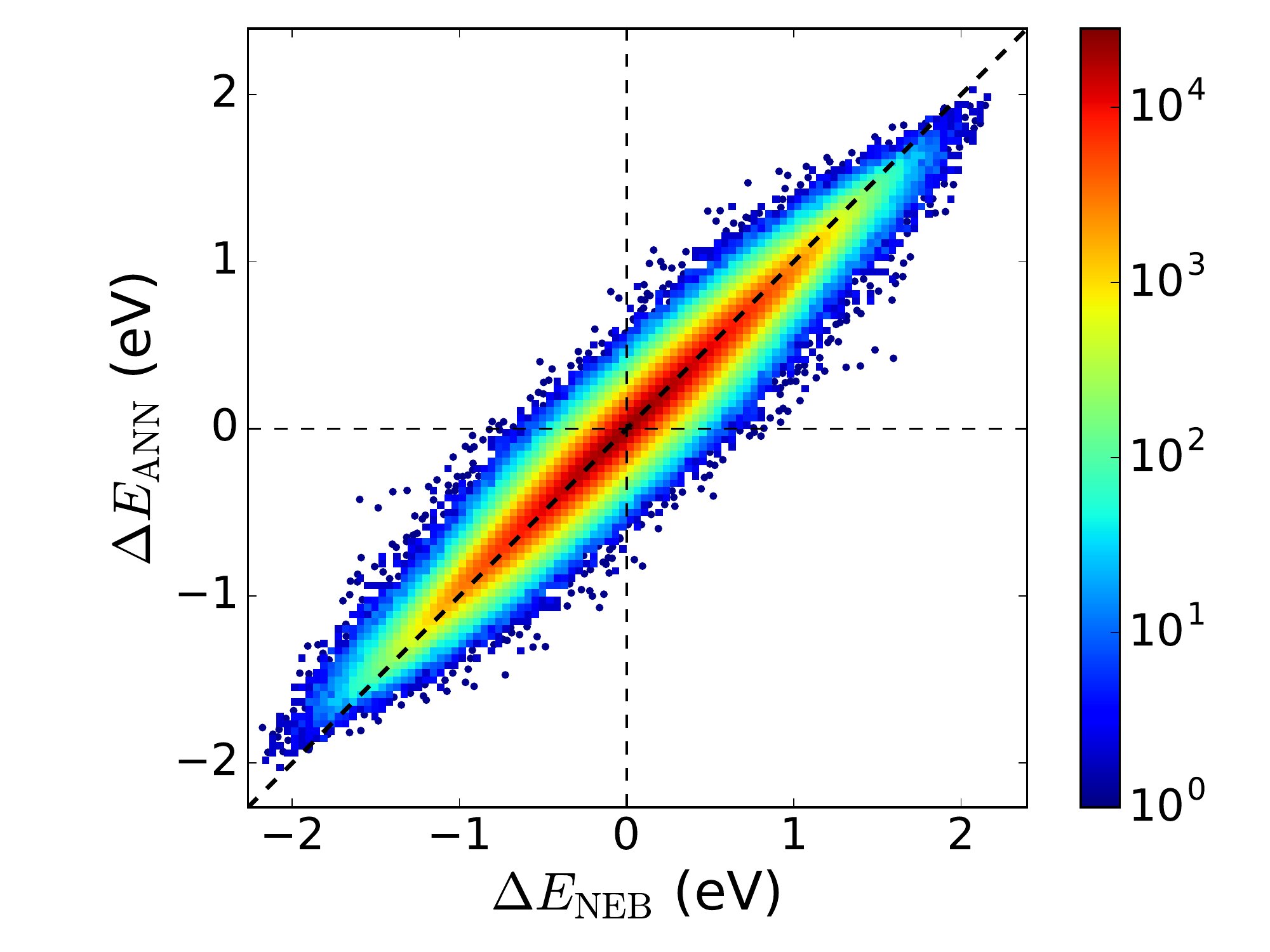}
      \caption{The correlation between true and predicted differences between the
               final and the initial energy values ($E_\mathrm{fin}-E_\mathrm{ini}=\Delta E$) of every jump.
               Roughly~6.7\,\% of the correlation points fall in the top left and the bottom right
               quadrants, representing cases where the energetically favourable jumps are
               predicted to be unfavourable and vice versa.}
      \label{fig:delta_e}
    \end{figure}

    Even though KMC does not have an explicit total potential energy parameter,
    the information about relative potential energy change in a jump
    can be inferred from the forward and the reverse barrier for that jump:
    $\Delta E = E_\mathrm{fin} - E_\mathrm{ini} = E_\mathrm{m}-E_\mathrm{m,reverse}$
    (see figure~\ref{fig:barrier}). For a KMC model to produce thermodynamically
    correct behaviour tending towards potential energy minima, these $\Delta E$
    values must be sufficiently accurate. We examined how accurately our machine learning model
    reproduces this thermodynamical information;
    fig.~\ref{fig:delta_e} shows the comparisons between the predicted $\Delta E$ and
    the values given by NEB. The overall correlation of the
    $\Delta E$ values is good, even though the model was fitted to
    the migration energy barriers only, and not the energy differences. We note, however,
    that in the region of barriers with near-zero values, the prediction of
    the $\Delta E$ values is less reliable. However, these barriers describe mainly
    unstable configurations that can be encountered due to limitations of the rigid
    lattice approximation. Since the barriers are low, they are not expected to
    affect the overall thermodynamic behaviour of the system.

    \subsection{KMC simulations}
    \label{subseq:kmcresults}
    
    We performed a number of KMC simulations to verify the applicability
    of the developed ANN for simulations of surface diffusion processes.
    We chose three typical scenarios of surface evolution driven by the
    surface energy minimisation principle. These are \emph{(i)} the flattening
    of a Cu nanotip on surfaces with different crystallographic orientation;
    \emph{(ii)} equilibrium shape of a Cu nanoparticle reached after relaxation of
    a particles with the different initial shape; \emph{(iii)} stability of Cu
    nanowires. All these processes can contribute to evolution of a rough
    surface with various surface features, which can be observed on surfaces
    subject to high electric fields. Since in the current study we aim to
    validate the proposed model, we focus on the processes under equilibrium
    condition (no electric field effects are yet taken into account). It
    is important to show that the model is stable and predicts physically
    reasonable behaviour of surfaces.
    
    \subsubsection{Nanotip flattening}
    \label{subsubsec:flattening}
\begin{figure}
        \centering
        \includegraphics[width=\linewidth]{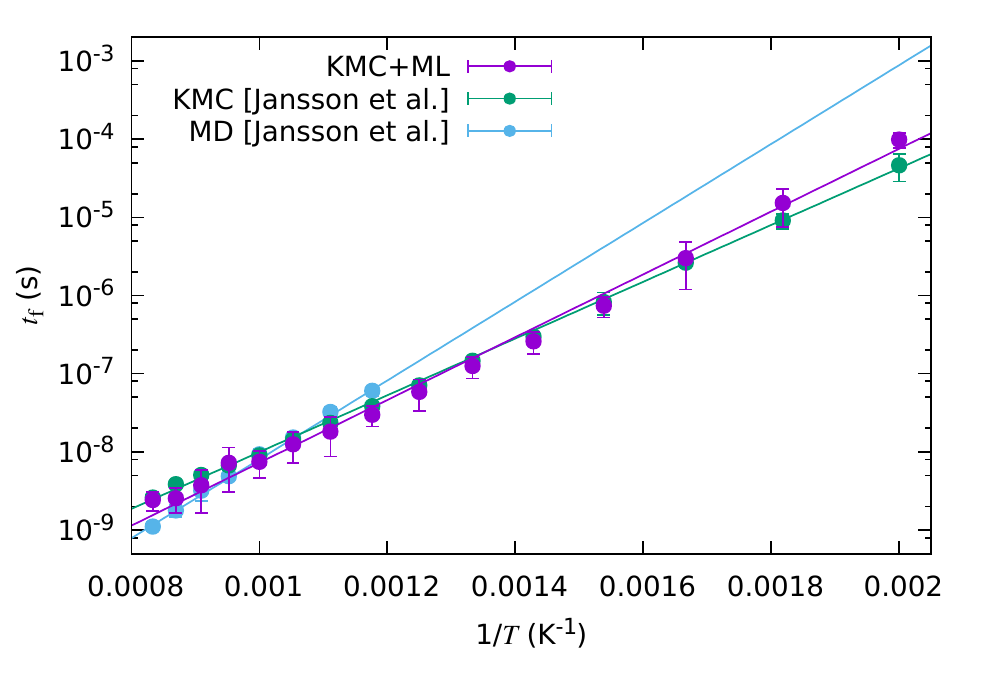}
        \caption{Flattening times of a 12 monolayer nanotip on the \hkl{110} surface at different temperatures.
                 The KMC and MD results are by Jansson et al.~\cite{jansson2016long}, and
                 KMC with machine learning (KMC+ML) refers to this work. The solid lines are fits of
                 eq.~\eqref{eq:flattening} to the data points.}
        \label{fig:fitting}
      \end{figure}

      A 12 monolayer nanotip flattening on \hkl{100}, \hkl{110}, and
      \hkl{111} surfaces was simulated with KMC using the trained
      combination MLP as a barrier predictor. The attempt frequency
      was fitted to the \hkl{110} case at different temperatures ranging from 850\,K
      to 1200\,K. For each temperature we performed 20 statistically different
      simulations. In some simulations at high temperatures, the surface started
      to self-roughen. We excluded these cases from the statistical averaging,
      since it was not clear how to account for the evolution of the height of
      the tip. See section~\ref{subsubsec:110_instability} for more detail.
      Hence the final number of cases used for statistical averaging were as follows: 18~(1000\,K),
      20~(1050\,K), 13~(1100\,K), 16~(1150\,K), 6~(1200\,K), and full~20
      at all lower temperatures.
      The attempt frequency value $\nu=2.81\times10^{14}$\,s$^{-1}$
      produced the best fit for the flattening time $t_\mathrm{f}$, comparing to the MD results
      ranging from 850\,K to 1200\,K, reported in ref.~\cite{jansson2016long} (see fig.~\ref{fig:fitting}).
      As the flattening times span over many orders of magnitude, the $\nu$
      parameter was fitted to minimise
      \begin{equation*}
      \mathrm{loss} = \left(\log t_\mathrm{f,ANN} - \log t_\mathrm{f,MD}\right)^2
      \end{equation*}
      The comparison shows the full parameterization to follow the trend observed in the MD
      simulations somewhat closer than the 4D parameterization, even though the difference is not drastic.
      
\begin{figure*}
        \centering
        \begin{subfigure}{0.32\linewidth}
          \captionsetup{labelformat=empty}
          \caption{\hkl{100} surface}
          \centering
          \begin{subfigure}{\linewidth}
            \centering
            \includegraphics[width=\linewidth]{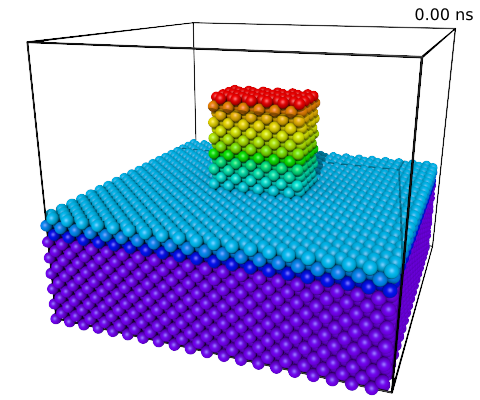}
          \end{subfigure}
          \begin{subfigure}{\linewidth}
            \centering
            \includegraphics[width=\linewidth]{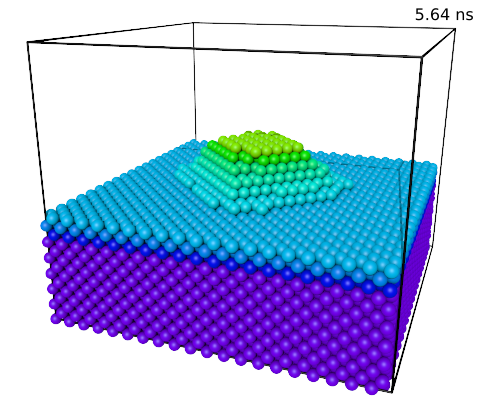}
          \end{subfigure}
        \end{subfigure}
        \vline
        \begin{subfigure}{0.32\linewidth}
          \captionsetup{labelformat=empty}
          \caption{\hkl{110} surface}
          \centering
          \begin{subfigure}{\linewidth}
            \centering
            \includegraphics[width=\linewidth]{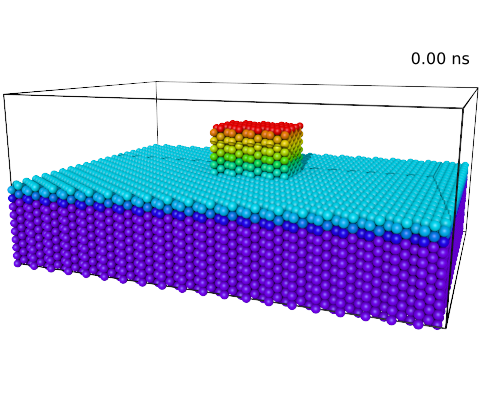}
          \end{subfigure}
          \begin{subfigure}{\linewidth}
            \centering
            \includegraphics[width=\linewidth]{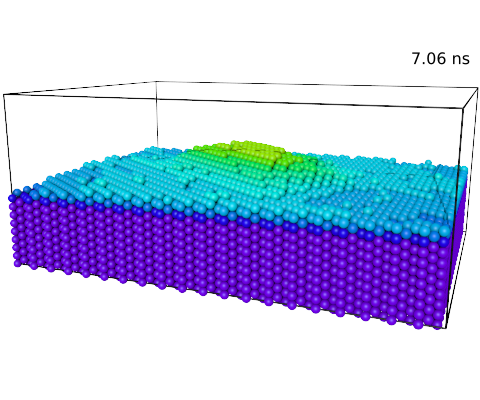}
          \end{subfigure}
        \end{subfigure}
        \vline
        \begin{subfigure}{0.32\linewidth}
          \captionsetup{labelformat=empty}
          \caption{\hkl{111} surface}
          \centering
          \begin{subfigure}{\linewidth}
            \centering
            \includegraphics[width=\linewidth]{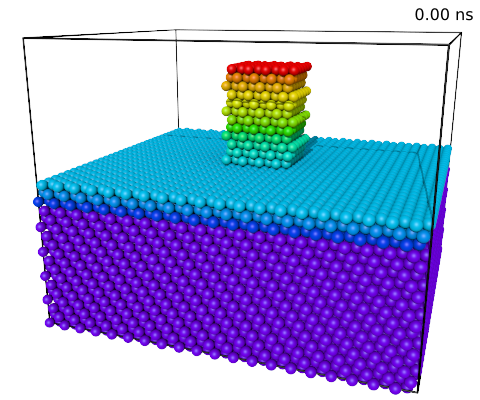}
          \end{subfigure}
          \begin{subfigure}{\linewidth}
            \centering
            \includegraphics[width=\linewidth]{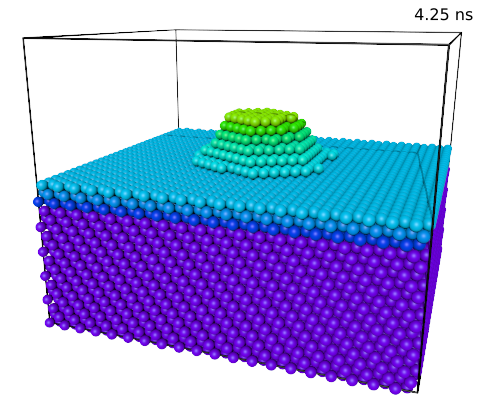}
          \end{subfigure}
        \end{subfigure}
        \caption{Examples of the initial and final frames of the cuboid nanotip flattening simulations
                 on the  \hkl{100}, \hkl{110} and \hkl{111} surfaces at 1000\,K.
                 The simulation was stopped
                 when the tip had reached half of its original height, replicating
                 the experiment in ref.~\cite{jansson2016long}.
                 The boundary conditions in both horizontal directions were periodic.
                 Colour coding (available online)
                 is according to the height coordinate of the atoms.}
        \label{fig:110_tip}
      \end{figure*}

      Arising from the event rate formula~\eqref{eq:KMC_rate}, the mean residence time $t_0$ and
      the mean migration barrier $E_\mathrm{a}$ can be obtained by fitting
      \begin{equation}
        \label{eq:flattening}
        t_\mathrm{f} = t_0 \exp\left(\frac{E_\mathrm{a}}{k_\mathrm{B}T}\right)
      \end{equation}
      to the flattening time data over a range of temperatures. The fitting
      parameters for the ANN parameterized KMC model, as well as for the KMC
      and the MD results from ref.~\cite{jansson2016long} are tabulated in table~\ref{tab:fitting}.      
      \begin{table}
        \caption{Parameters of eq.~\eqref{eq:flattening} fitted to the flattening times of a nanotip
                 on the \hkl{110} surface with three different models.}
        \centering
        \begin{tabular}{ccc}
          \toprule
          Model                      & $t_0$ (s)            & $E_\mathrm{a}$ (eV) \\
          \midrule
          KMC+ML                     & $7.01\cdot 10^{-13}$ & 0.80                \\
          KMC~\cite{jansson2016long} & $2.34\cdot10^{-12}$  & 0.72                \\
          MD~\cite{jansson2016long}  & $7.33\cdot 10^{-14}$ & 1.00                \\
          \bottomrule
        \end{tabular}
        \label{tab:fitting}
      \end{table}      
      With the fitted attempt frequency, the flattening of the tips on all
      three surfaces was simulated at 1000\,K. The mean results from 20 simulations
      are tabulated in table~\ref{tab:flattening}. In both tables we see much better agreement between the results
      obtained in the current work and those obtained with MD simulations.
      Fig.~\ref{fig:110_tip} shows the initial and the typical final configurations of these
      simulations.
      \begin{table}
        \caption{The mean flattening times of 12 monolayer, 576 atom cuboid nanotips
                 at 1000\,K on different surfaces.
                 KMC+ML refers to this work. The error estimates for the KMC+ML data are the standard deviations
                 of 20 simulations with different seeds in the case of the \hkl{100} and the \hkl{111}
                 surfaces, and standard deviations of 18 simulations in the case of \hkl{110}
                 surface---two simulations out of 20 resulted in the destabilisation of the surface.
                 See section~\ref{subsubsec:110_instability} for details.
                 Note that as the $\nu$ parameter is fitted on simulations on the \hkl{110} surface across a temperature
                 range from 850\,K to 1200\,K, the agreement to the MD result is not perfect
                 at the single data point of 1000\,K.}
        \centering
        \begin{tabular}{crrr}
          \toprule
          Surface & KMC+ML (ns)    & KMC~\cite{jansson2016long} (ns) & MD~\cite{jansson2016long} (ns) \\
          \midrule
          \hkl{100} & $5.2 \pm 0.6$ & $31.0 \pm 6.61$                 & $1.62 \pm 0.60$                \\
          \hkl{110} & $7.5 \pm 0.7$ & $9.25 \pm 1.10$                 & $9.29 \pm 1.44$                \\
          \hkl{111} & $4.8 \pm 0.9$ & $18.8 \pm 0.96$                 & $6.01 \pm 1.48$                \\
          \bottomrule
        \end{tabular}
        \label{tab:flattening}
      \end{table}
      
    \subsubsection{Instability of the \{1\,1\,0\} surface}
    \label{subsubsec:110_instability}

\begin{figure}
        \centering
        \includegraphics[width=0.9\linewidth]{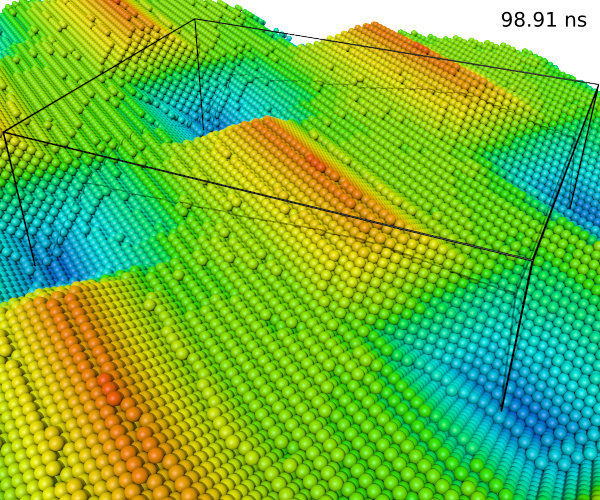}
        \caption{A valley (connected through periodic
                 boundaries) develops in some simulations at 1000\,K or higher
                 temperatures. The system is essentially stuck
                 in this configuration, with the valley growing deeper
                 with atoms accumulating from the bottom to the top of
                 the ridge as the simulation progresses. Colour coding
                 is according to the height coordinate of atoms. The simulation
                 box borders are shown in black; the box is replicated to
                 illustrate the shape of the valley.}
        \label{fig:unstable_110}
      \end{figure}

\begin{figure}
        \centering
        \includegraphics[width=\linewidth]{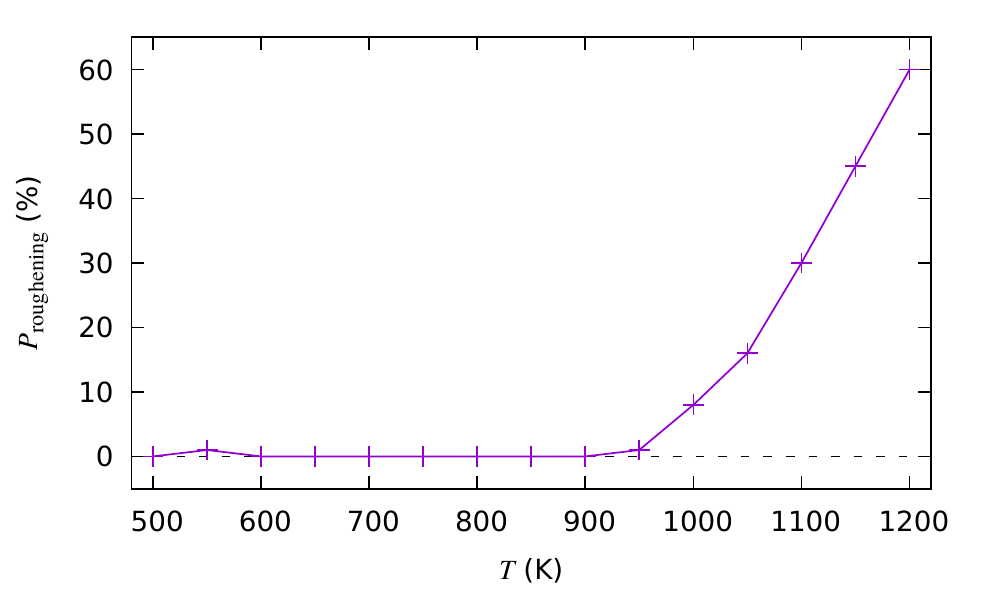}
        \caption{Probability for the \hkl{110} surface roughening before
                 the cuboid nanotip flattens, at different temperatures.
                 100 simulations were run at each temperature.}
        \label{fig:roughening}
      \end{figure}

      At temperatures $T\geq950$\,K, the \hkl{110} sometimes becomes
      unstable before the nanotip has time to flatten. In those cases, the surface near the nanotip
      starts to deplete, growing larger and larger \hkl{100} and \hkl{111} facets
      (see fig.~\ref{fig:unstable_110}).
      The atoms accumulate on top of the nanotip, which thus cannot be expected
      to flatten. The probability of destabilisation increases with
      temperature: see fig.~\ref{fig:roughening} for the probability of
      roughening as a function of temperature. Additional 100 simulations on the
      \hkl{110} surface were run for the statistics in that figure, with a slightly
      different settings for the simulation box: the horizontal dimensions were
      $118\times130$\,\AA$^2$, and surface was oriented to align
      atomic ridges of the \hkl{110} surface with the one of the sides of the simulation box.
    
    \subsubsection{Nanoparticle relaxation}
    \label{subsubsec:nanoparticles}

\begin{figure*}
        \centering
        \begin{subfigure}{0.19\linewidth}
          \centering
          \captionsetup{labelformat=empty}
          \caption{Sphere}
          \begin{subfigure}{\linewidth}
            \centering
            \includegraphics[width=0.4\linewidth]{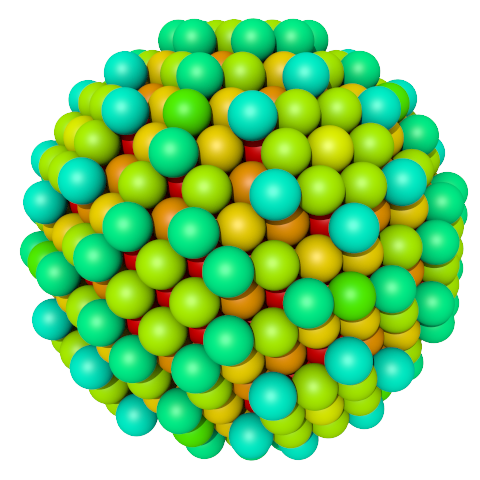}
            \raisebox{0.2\linewidth}{$\rightarrow$}
            \includegraphics[width=0.4\linewidth]{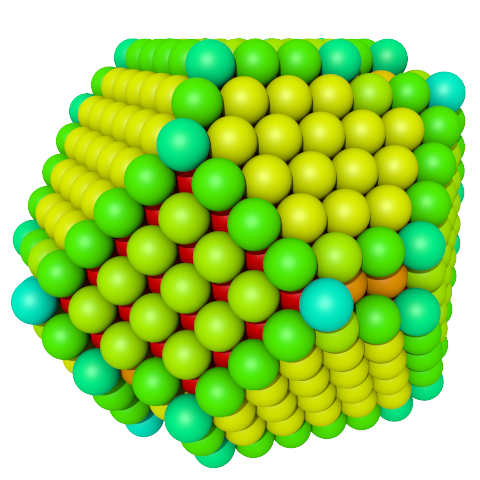}
          \end{subfigure}
          \captionsetup{labelformat=empty}
          \caption{887 atoms}
          \begin{subfigure}{\linewidth}
            \centering
            \includegraphics[width=0.4\linewidth]{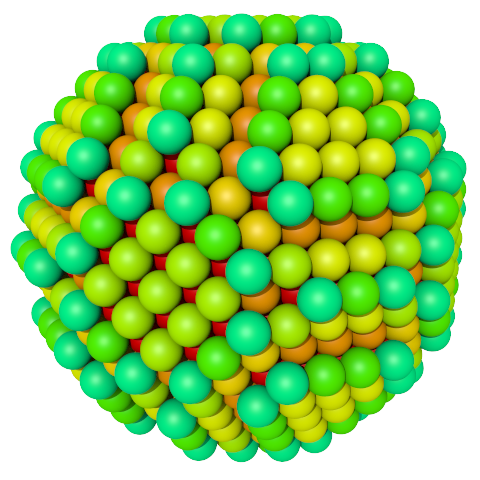}
            \raisebox{0.2\linewidth}{$\rightarrow$}
            \includegraphics[width=0.4\linewidth]{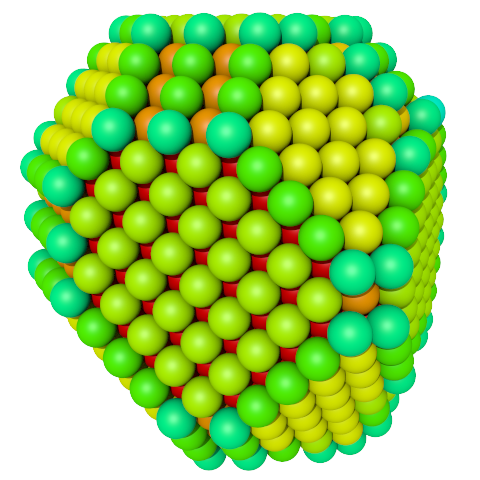}
          \end{subfigure}
          \captionsetup{labelformat=empty}
          \caption{1409 atoms}
          \begin{subfigure}{\linewidth}
            \centering
            \includegraphics[width=0.4\linewidth]{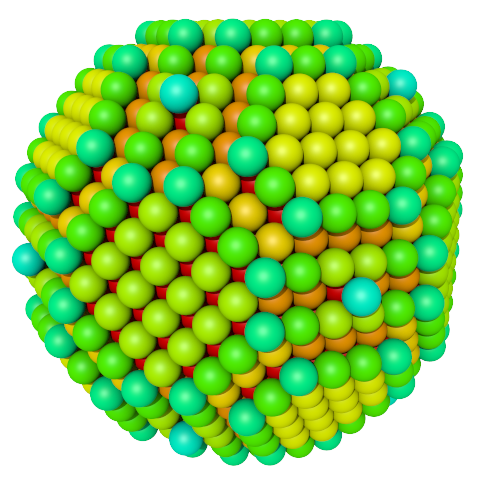}
            \raisebox{0.2\linewidth}{$\rightarrow$}
            \includegraphics[width=0.4\linewidth]{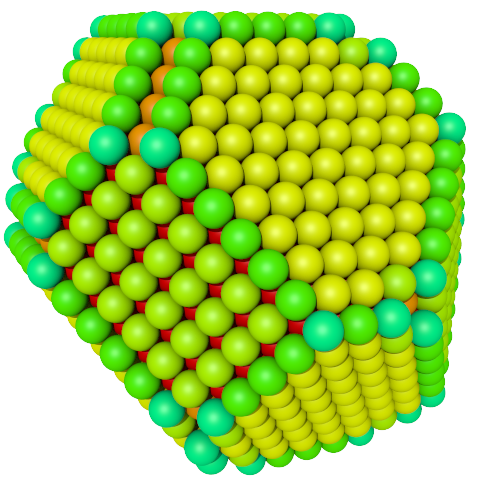}
          \end{subfigure}
          \captionsetup{labelformat=empty}
          \caption{2093 atoms}
        \end{subfigure}
        \vline
        \begin{subfigure}{0.19\linewidth}
          \centering
          \captionsetup{labelformat=empty}
          \caption{Octahedron}
          \begin{subfigure}{\linewidth}
            \centering
            \includegraphics[width=0.4\linewidth]{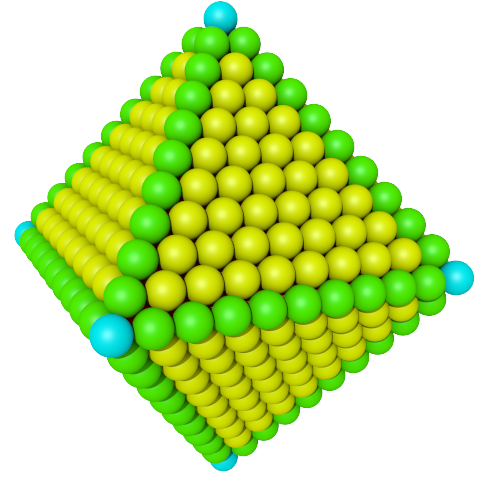}
            \raisebox{0.2\linewidth}{$\rightarrow$}
            \includegraphics[width=0.4\linewidth]{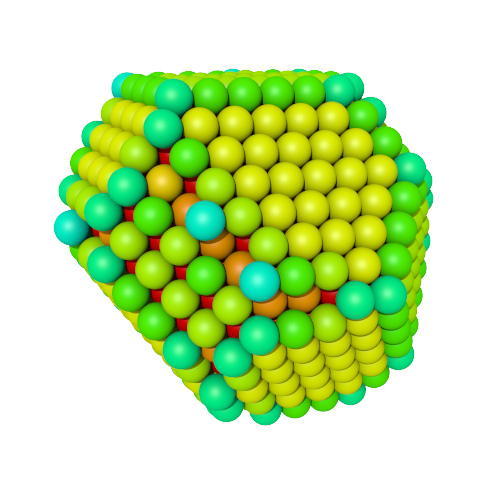}
          \end{subfigure}
          \captionsetup{labelformat=empty}
          \caption{891 atoms}
          \begin{subfigure}{\linewidth}
            \centering
            \includegraphics[width=0.4\linewidth]{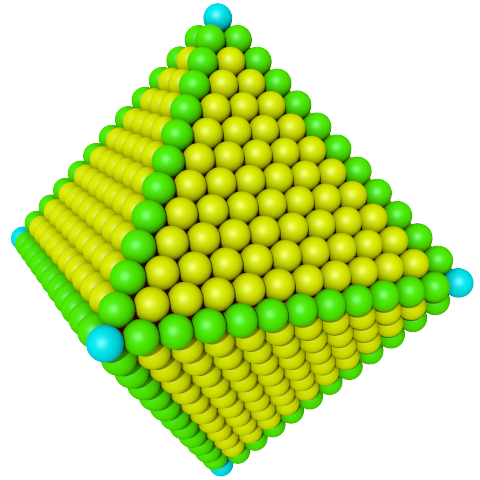}
            \raisebox{0.2\linewidth}{$\rightarrow$}
            \includegraphics[width=0.4\linewidth]{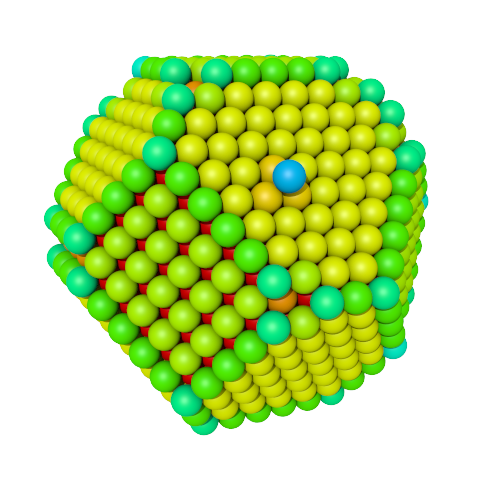}
          \end{subfigure}
          \captionsetup{labelformat=empty}
          \caption{1469 atoms}
          \begin{subfigure}{\linewidth}
            \centering
            \includegraphics[width=0.4\linewidth]{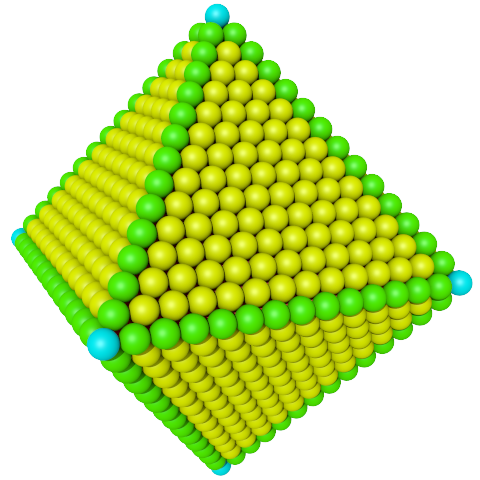}
            \raisebox{0.2\linewidth}{$\rightarrow$}
            \includegraphics[width=0.4\linewidth]{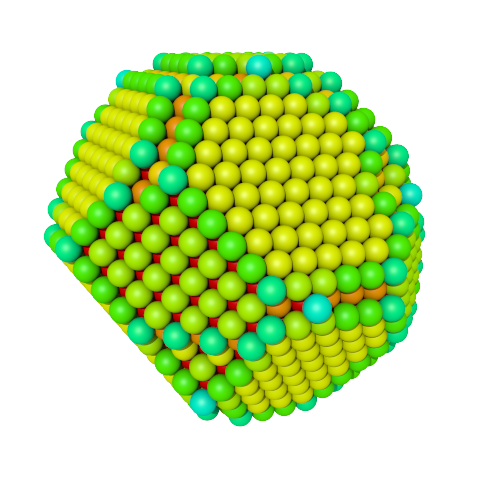}
          \end{subfigure}
          \captionsetup{labelformat=empty}
          \caption{2255 atoms}
        \end{subfigure}
        \vline
        \begin{subfigure}{0.19\linewidth}
          \centering
          \captionsetup{labelformat=empty}
          \caption{Cube}
          \begin{subfigure}{\linewidth}
            \centering
            \includegraphics[width=0.4\linewidth]{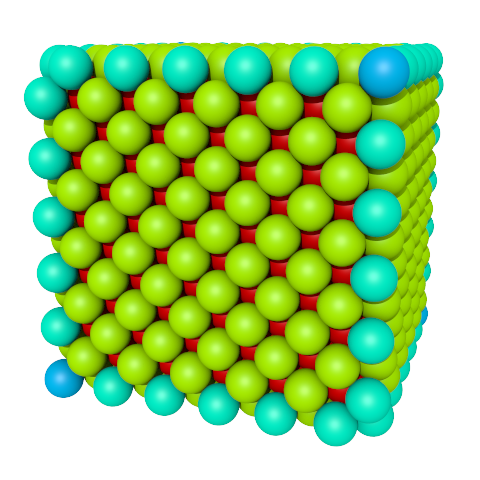}
            \raisebox{0.2\linewidth}{$\rightarrow$}
            \includegraphics[width=0.4\linewidth]{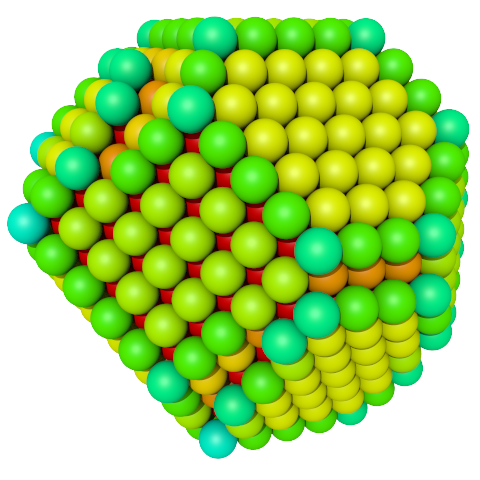}
          \end{subfigure}
          \captionsetup{labelformat=empty}
          \caption{864 atoms}
          \begin{subfigure}{\linewidth}
            \centering
            \includegraphics[width=0.4\linewidth]{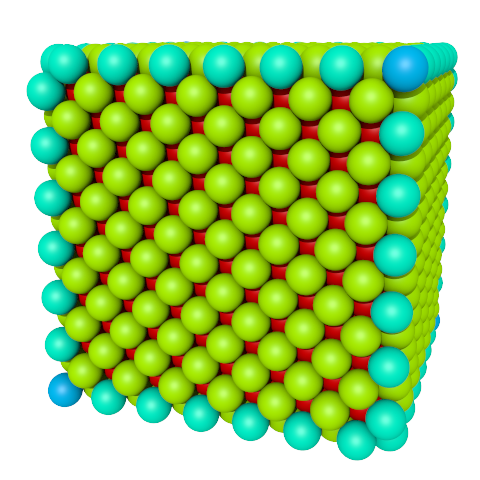}
            \raisebox{0.2\linewidth}{$\rightarrow$}
            \includegraphics[width=0.4\linewidth]{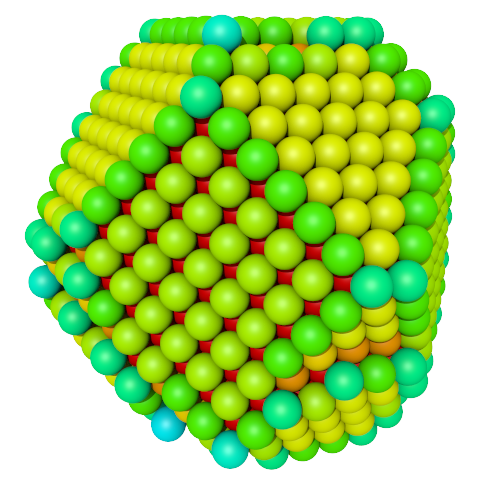}
          \end{subfigure}
          \captionsetup{labelformat=empty}
          \caption{1372 atoms}
          \begin{subfigure}{\linewidth}
            \centering
            \includegraphics[width=0.4\linewidth]{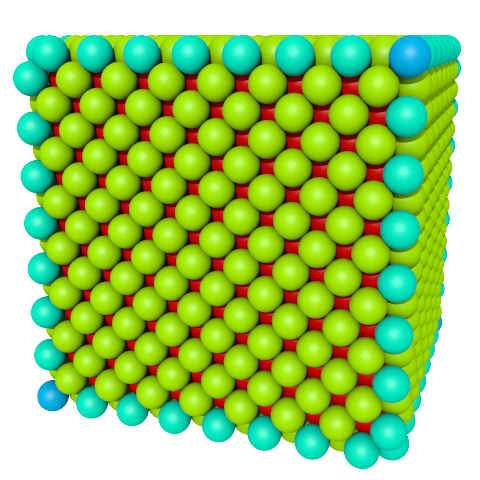}
            \raisebox{0.2\linewidth}{$\rightarrow$}
            \includegraphics[width=0.4\linewidth]{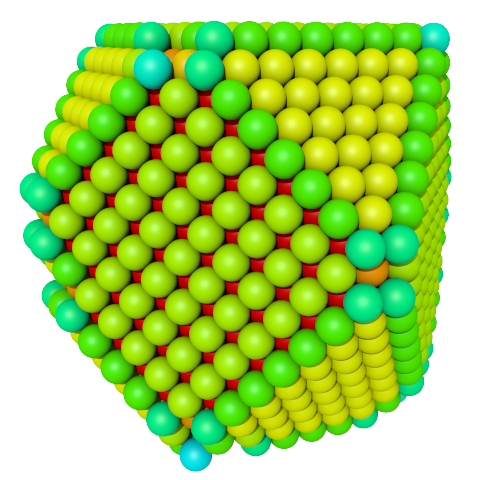}
          \end{subfigure}
          \captionsetup{labelformat=empty}
          \caption{2048 atoms}
        \end{subfigure}
        \vline
        \begin{subfigure}{0.19\linewidth}
          \centering
          \captionsetup{labelformat=empty}
          \caption{Cuboctahedron}
          \begin{subfigure}{\linewidth}
            \centering
            \includegraphics[width=0.4\linewidth]{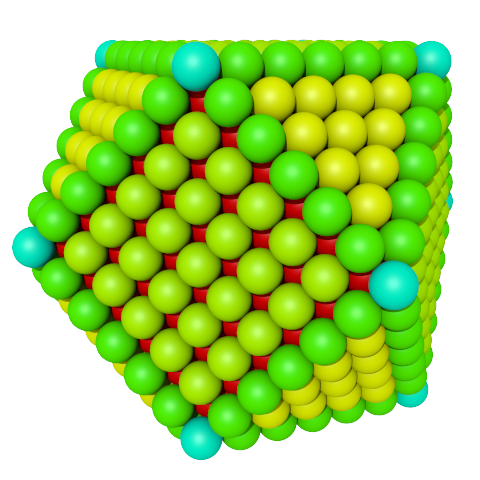}
            \raisebox{0.2\linewidth}{$\rightarrow$}
            \includegraphics[width=0.4\linewidth]{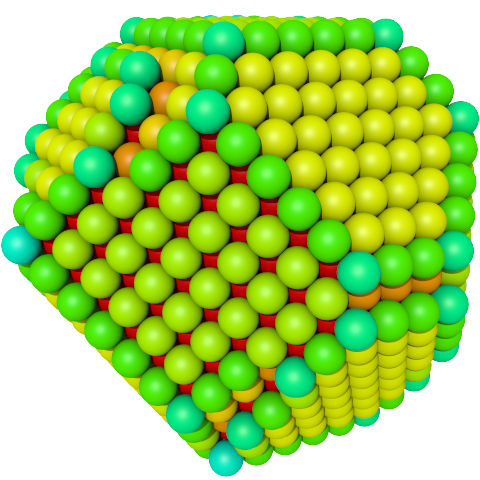}
          \end{subfigure}
          \captionsetup{labelformat=empty}
          \caption{923 atoms}
          \begin{subfigure}{\linewidth}
            \centering
            \includegraphics[width=0.4\linewidth]{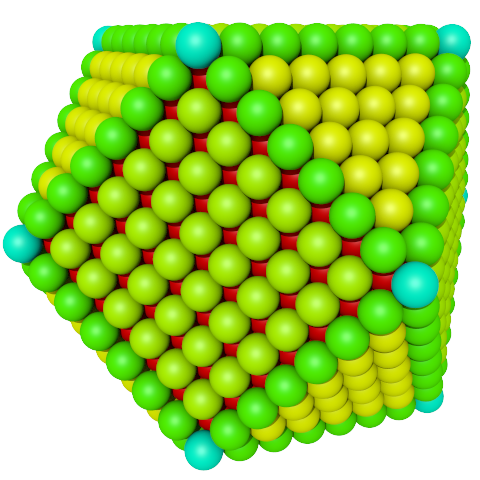}
            \raisebox{0.2\linewidth}{$\rightarrow$}
            \includegraphics[width=0.4\linewidth]{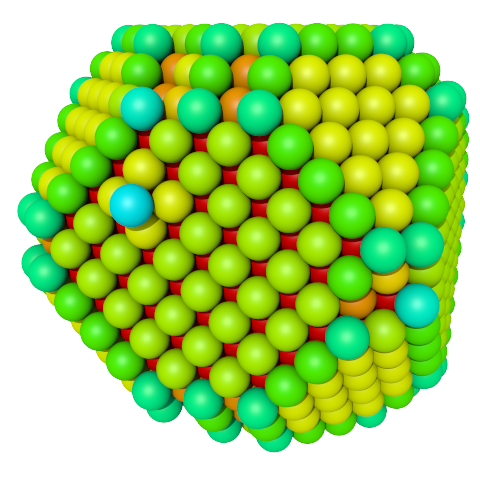}
          \end{subfigure}
          \captionsetup{labelformat=empty}
          \caption{1415 atoms}
          \begin{subfigure}{\linewidth}
            \centering
            \includegraphics[width=0.4\linewidth]{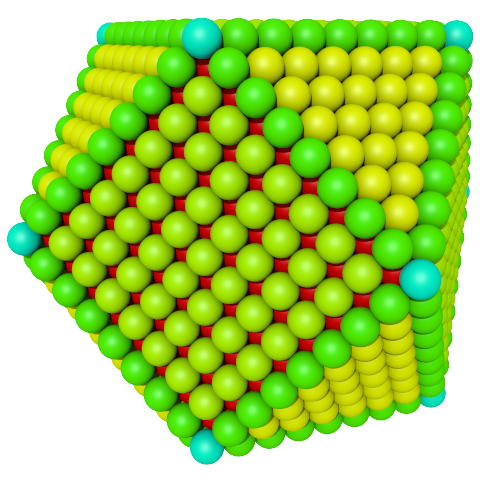}
            \raisebox{0.2\linewidth}{$\rightarrow$}
            \includegraphics[width=0.4\linewidth]{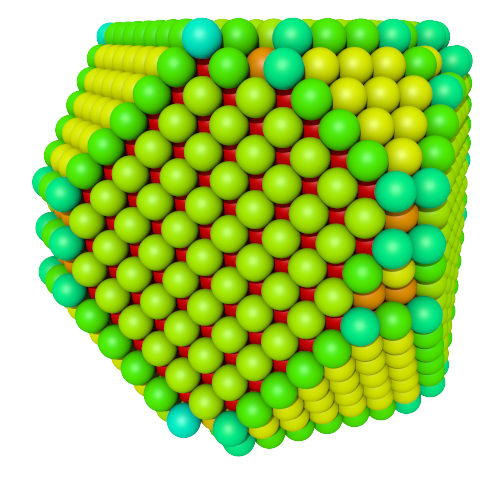}
          \end{subfigure}
          \captionsetup{labelformat=empty}
          \caption{2057 atoms}
        \end{subfigure}
        \vline
        \begin{subfigure}{0.19\linewidth}
          \centering
          \captionsetup{labelformat=empty}
          \caption{Wulff construction}
          \begin{subfigure}{\linewidth}
            \centering
            \includegraphics[width=0.4\linewidth]{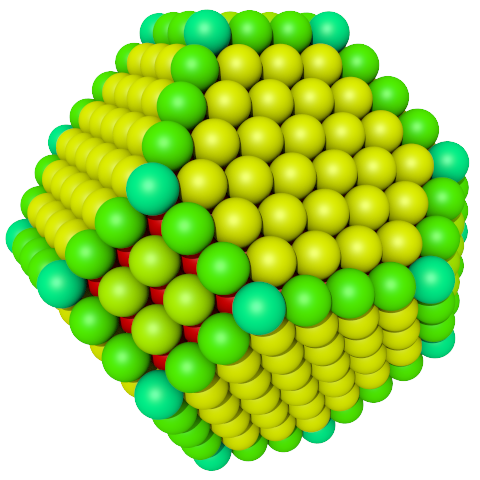}
            \raisebox{0.2\linewidth}{$\rightarrow$}
            \includegraphics[width=0.4\linewidth]{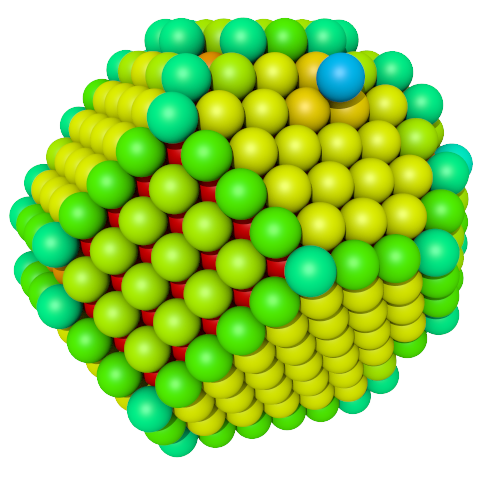}
          \end{subfigure}
          \captionsetup{labelformat=empty}
          \caption{807 atoms}
          \begin{subfigure}{\linewidth}
            \centering
            \includegraphics[width=0.4\linewidth]{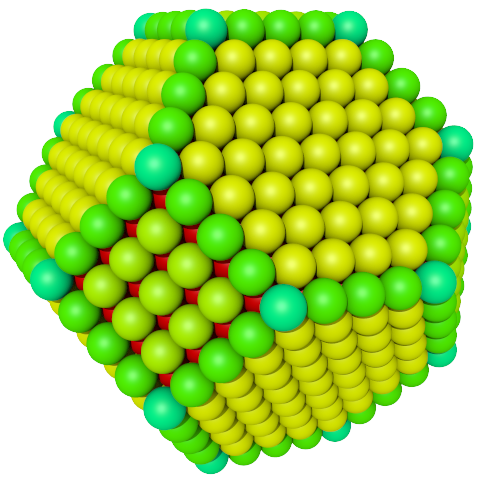}
            \raisebox{0.2\linewidth}{$\rightarrow$}
            \includegraphics[width=0.4\linewidth]{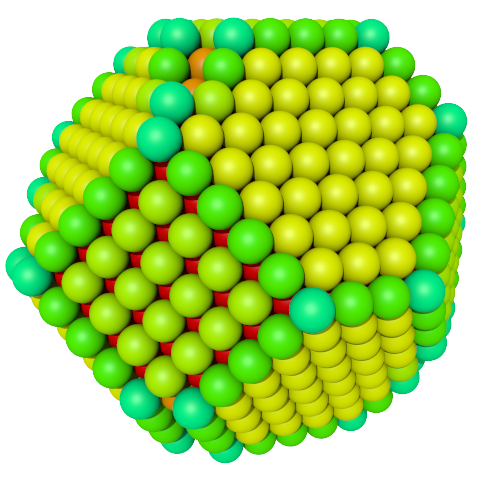}
          \end{subfigure}
          \captionsetup{labelformat=empty}
          \caption{1289 atoms}
          \begin{subfigure}{\linewidth}
            \centering
            \includegraphics[width=0.4\linewidth]{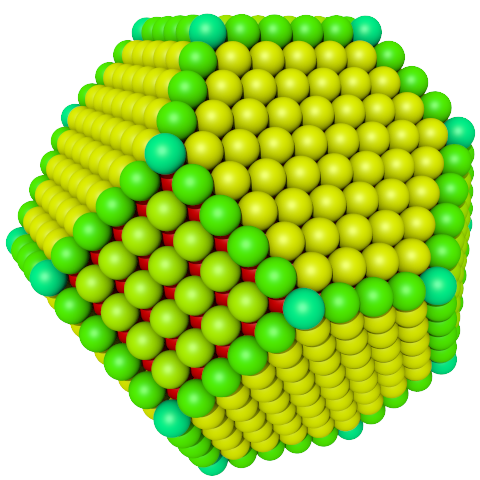}
            \raisebox{0.2\linewidth}{$\rightarrow$}
            \includegraphics[width=0.4\linewidth]{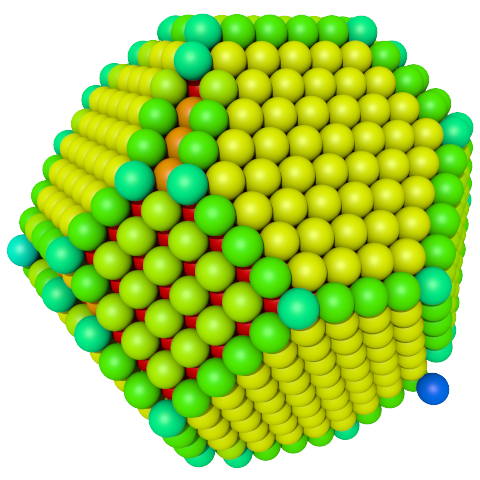}
          \end{subfigure}
          \captionsetup{labelformat=empty}
          \caption{1925 atoms}
        \end{subfigure}
        \caption{Examples of nanoparticle shapes before and after 1--4 microseconds
                 of KMC simulation at 1000\,K. Regardless of the initial shape,
                 the particles relax to the same final shape that is very
                 close to the minimum energy Wulff construction.} 
        \label{fig:final_particles}
      \end{figure*}

      Fig.~\ref{fig:final_particles} shows examples the shapes of the
      nanoparticles before and after KMC relaxation. Each simulation was performed at 1000\,K,
      each nanoparticle was evolving for about 1--4\,\textmu s, depending on the size.
      The figure is organized in panels separated by the thin vertical lines. Each panel
      is named according to the initial shape of the nanoparticles. Each panel shows three
      different size nanoparticles of the same initial shape, which are shown to the left
      of the panel and complemented by the final shape of the same nanoparticle after annealing on the right.
      The last panel shows the evolution starting from the minimal surface energy case, which is
      given by the Wulff construction, as generated by the Atomic Simulation Environment (ASE)~\cite{ase-paper}.
      At finite temperature, the shape should be expected to fluctuate near the
      minimum energy configuration, if the model produces correct surface energies.
      Since all particles eventually relaxed into shapes close to the one given by the
      Wulff construction, we conclude that our ANN model captures the surface energies
      correctly through the migration barrier information, and hence the thermodynamical
      evolution of the system given by the model is reliable.

    \subsubsection{Nanowire stability}
    \label{subsubsec:nanowires}
 
      We simulated Cu nanowires of 0.5\,nm radius at 1000\,K. Thin nanowires melt at temperatures
      considerably lower than bulk Cu; Granberg et al. found Cu nanowires of diameter 1.5\,nm to melt
      at 1000\,K, using a similar MC/MD-CEM potential for which the bulk melting points was 1656.72\,K~\cite{granberg2014investigation}.
      Furthermore, Cu nanowires are known to fragment by Rayleigh
      instability at much lower temperatures than the melting point~\cite{toimil2004fragmentation,li2008thermal}.
      Nevertheless, we chose a small wire radius and a temperature of 1000\,K to accelerate the simulation,
      with the primary goal of analyzing the performance of the current parameterization of KMC simulations of
      Rayleigh instability in thin wires as well as the mechanisms of such instability.
      
\begin{figure}
        \centering
        \includegraphics[width=\linewidth]{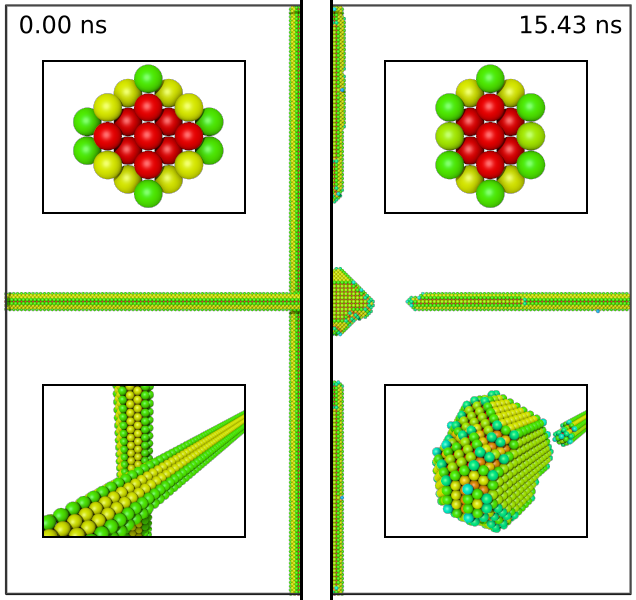}
        \caption{Evolution of a system of two crossing \hkl<110> nanowires. Left side of the figure shows the
                 initial configuration, and the right side shows the configuration after both wires have split
                 around the central junction. Top insets show the cross-section of a single wire, and bottom
                 insets show the central junction.}
        \label{fig:wires_break}
      \end{figure}

      In fig.~\ref{fig:wires_break} we show examples of the \hkl<110> nanowire junction simulations.
      In these simulations, atoms diffuse along
      the wires towards the central junction, tending to minimise the surface
      area of the structure. The sharp dips at the junction are filled up by
      the atoms from nearby regions of the intersecting nanowires. When a stable
      channel to deliver the atoms from the wire to the knob at the junction is
      established, the atoms move from large surface-to-ratio areas at a nanowire
      to the smaller surface-to-volume regions at the knob. This leads to thinning of the wires and eventually breaking near
      the center. The first breaking occurred right next to the junction in all of the 20 simulations.
      The mean time for the first breaking to occur in a system of crossing nanowires
      was $8 \pm 4$\,ns. Individual nanowires were much more stable, taking $250 \pm 130$\,ns
      to break by diffusion. Errors are standard deviations observed in 20 simulations with different random seeds.
      
      Fragmentation of Cu nanowire networks at the intersection points was observed experimentally by
      Mallikarjuna et al.~\cite{mallikarjuna2016photonic} and Oh et al.~\cite{oh2018efficient}.
      Similar behaviour was found in Au nanowires by Vigonski et al.~\cite{vigonski2017nanowire}
      both in experiments and in KMC simulations using the Kimocs code.

  \section{Discussions}
  \label{sec:discussions}
  
    The prediction performance of the ANNs in the set of
    the 3168 flat surface process barriers (fig.~\ref{fig:correlation_flat})
    is extremely good. This evidence supports
    the applicability of ANNs for surface migration
    barrier prediction and the validity of the 26D parameterization
    for describing the LAEs. The accuracy
    observed for the full~\textapprox11.7 million barrier set was
    considerably lower~(fig.~\ref{fig:FANN_full}).
    Some improvement was achieved by dividing the barrier data
    into three subsets by surface orientation---\hkl{100}, \hkl{110},
    and \hkl{111}---and training specialist networks to predict
    barriers on each surface. We propose that the improvement
    is due to the networks implicitly learning the effect of
    the different surface stresses on the migration barriers.
    Further accuracy was gained by training ANN ensembles
    to each barrier subset, and averaging the predicted barrier
    over them. Even with these
    additional techniques, an accuracy comparable to the
    flat surface case was not reached.

    The reason for the loss of accuracy in the full set is not entirely clear.
    The full set is three orders of magnitude larger than the set
    of flat surface barriers, with a more
    heterogeneous composition of the LAEs and a larger range of barrier values.
    The data itself seems also to be of reasonable
    quality, as the overall agreements with previous
    barrier sets (figs.~\ref{fig:compare_notether}
    and~\ref{fig:compare_tether}) are good.
    
    One possible explanation is, in addition to the different
    ranges, the different distributions of barrier values in the sets.
    The full set distribution (fig.~\ref{fig:full_histogram}) resembles
    half of a Gaussian distribution except for the large amount of zero barriers,
    while the 3000-barrier flat set distribution (fig.~\ref{fig:flat_histogram})
    has no such ``discontinuities''.
    This kind of discontinuity in the output value distribution
    may be one of the reasons for the difficulties in the training.
    To verify this, in future works it could be considered
    to assign some negative pseudo-barriers to the 0\,eV cases for
    the duration of the training to make the output distribution smoother.
    During KMC, negative barriers given by the ANN would be set back to zero.
    Another method to smoothen the distribution could be by introducing
    a finer classification scheme: in addition to splitting the data set
    into subsets of \hkl{100}, \hkl{110}, and \hkl{111} processes, further
    subdivisions could be made based on the jumping direction
    on the surface or on the stability of the LAE (e.g. how many neighbours
    the least bound atom in the LAE has).
    
    The complexity of the network architecture could be questioned as a potential
    culprit for the lack of accuracy---are single layer ANNs and cascade networks
    sophisticated enough to produce sufficient fitting?
    However, the networks used in this work are very similar to what has been found to
    work well in earlier studies on machine learning for barrier prediction.
    References~\cite{djurabekova2007artificial,castin2009prediction,castin2009modelling,
    castin2010calculation,castin2011modeling,pascuet2011stability,castin2012mobility,
    castin2017improved,messina2017introducing} use
    ANNs with only one hidden layer. Cascade networks are used in
    refs.~\cite{castin2008use} and~\cite{castin2014predicting}.
    The earlier rigid lattice variants of the method also use a similar descriptor:
    a vector of lattice sites encoded with integers according to the occupation state
    of each site. One of the differences in the
    earlier studies compared to this work is the size of the LAE---up to
    hundreds of atoms instead of the 26 used here. Expanding
    the LAE could thus provide an additional way to improve accuracy,
    although in ref.~\cite{sastry2005genetic}
    genetic programming was successfully applied to predict barriers
    using LAE with up to 2nn sites in the limited case of vacancy-assisted diffusion.
    
    Despite the apparent lower accuracy of the ANN in the full barrier set,
    it produced physically reasonable results for nanotip flattening,
    nanoparticle shape relaxation and simulations of nanowire instability. The
    attempt frequency value $2.81\times10^{14}$\,s$^{-1}$, obtained
    by fitting the flattening time of the nanotip on the \hkl{110} surface to the
    flattening time in MD, falls right within the range of the previously obtained values
    ($0.7\times10^{14}$~\cite{jansson2016long} and $3.1\times10^{14}$\,s$^{-1}$~\cite{baibuz2018migration}).
    The flattening times for the nanotips on the \hkl{100} and
    the \hkl{111} surfaces are also much closer to the MD results than with
    the KMC model proposed in ref.~\cite{jansson2016long} (see table~\ref{tab:flattening}),
    although the flattening mechanism is different than seen in MD.
    Due to limitations of the rigid lattice KMC, the reorientation of the crystal
    structure, which is commonly seen within the nanotip oriented in \hkl<100> direction
    \cite{jansson2016long,park2005shape,liang2005shape}, is not possible within this approach.
    
    As was mentioned in the results section, the thermodynamics of the Cu surface system appear
    to be well captured by our model, even though it was not explicitly trained to do so. Evidence towards this is given
    both by the prediction errors of the energy differences (see fig.~\ref{fig:delta_e}) and the relaxation
    of the nanoparticle shapes (fig.~\ref{fig:final_particles}).
    Only a small fraction of the jumps have their energy differences predicted
    in the opposite order.
    
    Regarding the instability of the \hkl{110} surface that was observed at $T>950$\,K,
    it is worth mentioning that the real Cu \hkl{110} surface is known to self-roughen
    at high temperatures. The lower bound
    for the roughening temperature $T_\mathrm{R}$ was experimentally estimated to be at~600\,\textdegree C (approx. 900\,K)
    initially by Mochrie~\cite{mochrie1987thermal} and later by
    Zeppenfeld et al.~\cite{zeppenfeld1989no}. Kern
    estimated $T_\mathrm{R}$ to be 1070\,K~\cite{kern1992thermal}.
    H\"akkinen et al. found a clear roughening transition
    to happen at~1000\,K in molecular dynamics simulations~\cite{hakkinen1993roughening}.
    
    In experiments, the roughening phenomenon is usually observed indirectly from changes
    in the diffraction of e.g. X-rays or He-atoms from the surface. In computational
    and theoretical models, the roughening is perceived as the proliferation of
    atomic steps, as the formation energy of the steps disappears upon
    reaching $T=T_\mathrm{R}$. In molecular dynamics simulations, roughening
    precedes \emph{premelting}, where the whole surface becomes covered by a thin liquid-like
    layer at a temperature below the bulk melting temperature~\cite{hakkinen1992computer}. While
    roughening---formation of atomic steps---could in principle
    be modelled in rigid lattice KMC, premelting certainly cannot.
    Further studies are required as to whether the behaviour observed in this
    work is a good description of the roughening phenomenon, or if the transition
    at 1000\,K is more of a coincidence.
    In any case, we advise caution when modelling the Cu \hkl{110} surface
    at elevated temperatures with this ANN KMC parameterization.
    
    As for the results for the stability of individual and crossing thin \hkl<110>
    nanowires, the model suggests that fragmentation occurs first near the junctions
    of two nanowires. Individual nanowires took approximately 250\,ns to break,
    while the first breaking occurred in approximately 8\,ns in the systems of two nanowires.
    This behaviour is similar to what was earlier shown for Au wires
    in both simulations and experiments~\cite{vigonski2017nanowire}: nanowire systems
    tend to break at the junction points earlier and at lower temperatures than
    individual nanowires. The Au and Cu systems are rather well comparable, since both
    metals are fcc materials with similar relative surface energies on the low index facets.
    Some experimental evidence for similar behaviour in Cu exists: Mallikarjuna et al.
    observed increased electrical resistance in Cu nanowire mesh if photonic
    welding was carried on excessively long~\cite{mallikarjuna2016photonic}. They attributed the
    loss of resistance to breakup of nanowire junctions, which they also saw in
    FE-SEM images. Oh et al. also saw broken Cu nanowire
    junctions after using very high current in Eddy current welding~\cite{oh2018efficient}, although
    they propose that breaking was due to oxidisation of the junction; in our model, oxygen is
    not present. Despite this limitation, the similarity of our results compared with these
    experiments and with previous results for Au wires, provides sufficient
    validation of the proposed model.
    
  \section{Conclusions}
  \label{sec:conclusions}

    We have developed an artificial neural network that can predict Cu surface migration
    energy barriers with sufficient accuracy and implemented it into a Kinetic Monte Carlo
    model. The parameterization is fully three-dimensional, i.e. it is applicable to
    arbitrarily rough surfaces. The Kinetic Monte Carlo model is able to accurately simulate the energy
    minimisation of Cu nanoclusters, the thermal flattening of small nanotips as well
    as the fragmentation of nanowires. Our model predicts the \hkl{110} surface
    to be unstable at temperatures above 1000\,K, very near to the known roughening
    temperature of this Cu surface. The exact mechanism of the instability may
    be different from actual roughening, so we advise some caution when using
    the model at high temperatures.
    
  \section*{Acknowledgements}
  
    J.\;Kimari was supported by a CERN K-contract and Academy
    of Finland (Grant No.\;313867).
    V.\;Jansson was supported by Academy of Finland (Grant No.\;285382).
    F.\;Djurabekova acknowledges gratefully the financial support of
    Academy of Finland (Grant No.\;269696) and MEPhI Academic Excellence
    Project (Contract No.\:02.a03.21.0005).
    E.\;Baibuz was supported by a CERN K-contract and the doctoral
    program MATRENA of the University of Helsinki.
    V.\;Zadin and S.\;Vigonski were supported by Estonian Research
    Council grant PUT 1372. Computing resources were provided
    by the Finnish IT Center for Science (CSC) and the Finnish
    Grid and Cloud Infrastructure (persistent
    identifier urn:nbn:fi:research-infras-2016072533).
    
  \section*{Data availability}
    
    The raw data required to reproduce these findings
    are available to download from \url{http://urn.fi/urn:nbn:fi:att:cde3019b-d849-41d2-a04a-9f7eb2972f5a}.
    The processed data required to reproduce these findings are
    available to download from \url{http://urn.fi/urn:nbn:fi:att:a4120c3b-9535-405d-b768-4a972bce0b1b}.
    
\section*{References}
\bibliography{jyri_kimari.bib}

\end{document}